\documentclass[aps,prfluids]{revtex4}
\usepackage{graphicx}
\usepackage{bm}
\usepackage{amsmath}
\usepackage{amssymb}
\usepackage[svgnames]{xcolor}
\usepackage{verbatim}
\usepackage{enumitem}
\usepackage{subfigure}
\newcommand{\avg}[1]{\langle #1 \rangle}
\newcommand{\refeq}[1]{(\ref{#1})}

\begin{document}
\title{A new self-consistent approach of quantum turbulence in superfluid helium}
%Quantum turbulence in superfluid helium: \\
%a fully-coupled local model}
%\subtitle{Do you have a subtitle?\\ If so, write it here}

\author{Luca Galantucci, Andrew W. Baggaley, Carlo F. Barenghi}
%\email{c.f.barenghi@ncl.ac.uk}
\affiliation{
Joint Quantum Centre Durham-Newcastle, 
School of Mathematics, Statistics and Physics, Newcastle University,
Newcastle upon Tyne, NE1 7RU, UK
}

%\author{and Giorgio Krstulovic}
%\affiliation{
%Universit\'e C\^ote d'Azur, Observatoire de la C\^ote d'Azur, CNRS, Laboratoire Lagrange, Nice, France
%}

%\author{Andrew W. Baggaley}
%\affiliation{
%Joint Quantum Centre Durham-Newcastle, 
%School of Mathematics, Statistics and Physics, Newcastle University,
%Newcastle upon Tyne, NE1 7RU, UK
%}
%\author{Carlo F. Barenghi}
%%\email{carlo.barenghi@newcastle.ac.uk}
%\affiliation{
%Joint Quantum Centre Durham-Newcastle, 
%School of Mathematics, Statistics and Physics, Newcastle University,
%Newcastle upon Tyne, NE1 7RU, UK
%}%

\author{and Giorgio Krstulovic}
\affiliation{
Universit\'e C\^ote d'Azur, Observatoire de la C\^ote d'Azur, CNRS, Laboratoire Lagrange, Nice, France
}

\begin{abstract}
We present the Fully cOUpled loCAl model of sUperfLuid Turbulence (FOUCAULT) that describes the dynamics of finite temperature superfluids. The superfluid component is
described by the vortex filament method while the normal fluid 
is governed by a modified Navier-Stokes equation.
The superfluid vortex lines and normal fluid components are 
fully coupled in a self-consistent manner by the friction
force, which induce local disturbances in the normal fluid
in the vicinity of vortex lines. 
The main focus of this work is the numerical scheme for 
distributing the friction force to the mesh points where the normal 
fluid is defined and for evaluating the velocity of the normal
fluid on the Lagrangian discretization points along the vortex lines.  
In particular, we show that if this numerical scheme 
is not careful enough, spurious results may occur.
The new scheme which we propose to overcome these difficulties 
is based on physical principles. Finally, we apply the new method
to the problem of the motion of a superfluid vortex ring in
a stationary normal fluid and in a turbulent normal fluid. 
\end{abstract}

\maketitle

%%%%%%%%%%%%%%%% INTRODUCTION  %%%%%%%%%%%%%%%%%%%%%%%%%%

\section{Introduction}
\label{intro}

Quantum turbulence 
\cite{vinen-niemela-2002,skrbek-sreenivasan-2012,
barenghi-skrbek-sreenivasan-2014} 
is the disordered motion of quantum fluids - fluids which are governed
by the laws of quantum mechanics rather than classical physics.
Examples of quantum fluids are atomic Bose-Einstein condensates 
\cite{pitaevskii-stringari-2003,primer2016},
the low temperature liquid phases of helium isotopes
$^3$He and $^4$He, polariton condensates,
and the interior of neutron stars. The underlying physics is the 
condensation of atoms which obey Bose-Einstein statistics.
In this work we focus on the superfluid phase of $^4$He, 
frequently referred to as helium-II or He-II for short. 
Helium~II can be described as an intimate mixture of two 
fluid components \cite{tisza-1938,landau-1941}: 
the viscous \textit{normal fluid}, which is similar to
ordinary viscous fluids obeying the classical
Navier-Stokes equation,
and the inviscid \textit{superfluid}, whose vorticity is  
confined to \textit{vortex lines} of atomic thickness
and quantised circulation (also referred to as quantised vortices
or superfluid vortices). At nonzero temperatures,
quantum turbulence thus manifests itself
as a disordered tangle of vortex lines interacting with either a laminar or
a turbulent normal fluid depending on the circumstances.

%\red{In the absence of vortex lines, the motion of helium~II
%is well described by \textit{Landau-Tisza's two-fluid theory}, which
%predicts striking effects such as second sound and thermal counterflow.
%However, most flows of current interest 
%contain vortex lines, either because they rotate or because the velocity
%is sufficiently large. Vortex lines scatter
%the thermal excitations which make up the normal fluid, and thus introduce
%a non-linear mutual friction force which couples normal fluid and superfluid.
%This work deals with this coupling force.}

Despite the two-fluid nature and the quantisation of superfluid vorticity, 
several surprising analogies between classical turbulence and quantum 
turbulence have been established in helium~II in the last years.
One example is the same temporal decay of vorticity 
\cite{stalp-skrbek-donnelly-1999} in unforced turbulence.
Another example is the energy spectrum of forced homogeneous
isotropic turbulence, which,
according to experiments
\cite{maurer-tabeling-1998,salort-etal-2010,barenghi-lvov-roche-2014},
numerical simulations
\cite{baggaley-laurie-barenghi-2012,baggaley-barenghi-2011,
araki-tsubota-nemirovskii-2002,kobayashi-tsubota-2005}
and theory
\cite{lvov-nazarenko-skrbek-2006,lvov-nazarenko-rudenko-2008},
displays the same Kolmogorov energy spectrum
(the distribution of kinetic energy over the length scales)
which is observed in classical turbulence.
However, at sufficiently small length scales 
(smaller than the average inter-vortex distance), or when
superfluid and normal fluid are driven thermally in opposite directions
\cite{vinen-1957a,barenghi-sergeev-baggaley-2017}, 
fundamental differences emerge between classical
and quantum turbulence
\cite{lamantia-duda-rotter-skrbek-2013b,lamantia-2017}.

The comparison between quantum and classical turbulence 
is therefore a current
topic of lively discussions in low temperature physics community.
Essential progress in the understanding of the underlying physics 
is provided by the recent development of new experimental techniques 
for the visualisation of superfluid helium flows, including
micron-sized tracers,  polymer particles 
\cite{zhang-vansciver-2005}, solid hydrogen/deuterium flakes 
\cite{bewley-lathrop-sreenivasan-2006,lamantia-chagovets-rotter-skrbek-2012} 
and, more recently, smaller (thus less intrusive) metastable 
helium molecules \cite{guo-etal-2010} which fluoresce when excited by
a laser.

Alongside experiments, numerical simulations have played an important
role in understanding the physics of quantum turbulence, from 
interpreting  experimental data to proposing new experiments.
However, most numerical simulations have determined the motion of
vortex lines as a function of prescribed normal fluid
profiles without taking into account the back-reaction 
of the vortex lines \cite{hanninen-baggaley-2014} onto the normal fluid. 
Only a small number of investigations have addressed this back-reaction,
but most efforts have suffered from some
shortcomings: they were restricted to two-dimensional channels 
\cite{galantucci-sciacca-barenghi-2015,galantucci-sciacca-barenghi-2017},
had too limited numerical resolution to address fully developed 
turbulence \cite{yui-etal-2018}, considered only
simple vortex configurations 
\cite{kivotides-barenghi-samuels-2000,idowu-willis-barenghi-samuels-2000},
were limited to decaying turbulence
\cite{kivotides-2015jltp}), or did not reach the statistically steady-state 
which is necessary to make direct comparison to classical turbulence.

This work presents a fully-coupled, three dimensional
numerical model which tackles
all these shortcomings and contains innovative features concerning 
the numerical architecture and the physical modelling of the 
interaction between normal fluid and vortex lines. 
We shall refer to our model as \textbf{FOUCAULT}, \textit{Fully-cOUpled loCAl model of sUperfLuid Turbulence}.
Compared to past studies, our model consists of an
efficiently parallelised pseudo-spectral code capable of distributing the 
calculation of the normal fluid velocity field and the temporal
evolution of the superfluid vortex tangle amongst 
distinct computational cluster nodes. This feature allows the resolution of a wider
range of flow length scales compared to the previous literature, spanning 
large {\it quasi-classical length scales} as well as smaller
{\it quantum length scales}.
This wide range is of fundamental importance for the numerical 
simulation of turbulent flows of helium~II with realistic experimental
parameters. 
%In addition, the inclusion of a normal fluid forcing scheme borrowed from classical turbulence allows in principle to achieve statistically stationary
%quantum turbulence.

From the physical point of view, FOUCAULT
includes a new approach to determine the friction force 
between superfluid vortices and the normal fluid, developed
from progress in the study of classical creeping flows \cite{kivotides-2018}, 
but modified in order to avoid the unphysical dependence of the 
friction on the numerical discretization on the vortex lines. 
In addition, the force between the vortices and the normal fluid 
(ideally a Dirac delta function centred on the vortex lines) is regularised
{\it exactly} employing a method
recently developed in classical turbulence 
\cite{gualtieri_picano_sardina_casciola_2015,GualtieriPREActiveParticlesPRE} 
for the consistent modelling of the two-way coupling between a viscous 
fluid and small active particles. This method
stems from the small-scale viscous diffusion of the normal fluid 
disturbances generated by the vortex motion, thus regularising 
the fluid response to vortex forcing in a physically consistent manner.
Our fully-coupled local model is therefore a significant 
improvement with respect to previous algorithms which
employed arbitrary numerical procedures for the distribution of the vortex 
forcing on the Eulerian computational grid of the
normal fluid \cite{kivotides-2011}. 

The organisation of the paper is the following.
In Section~\ref{sec:eqs} we describe 
the equations of motion of the superfluid, the normal fluid and 
the vortex lines with emphasis on the different 
existing models for the calculation of the friction force. 
In Section~\ref{sec:numerical}
we outline our numerical method, including details on the Vortex Filament 
Method, the Navier--Stokes solver,
interpolation schemes and the procedure employed for regularising 
the friction force consistently. Next, in Section~\ref{sec:examples}, we
%validate our algorithm and 
show two applications of our model.
Finally, in Section~\ref{sec:concl} 
we summarise the main points and outline future work. 
%of the disturbances generated in a viscous fluid by small, spherical 
%particles in relative motion with respect to the fluid itself. This modelling  
%
%in order to model the injection of vorticity in the normal fluid flow due to the relative   
%\vspace{1cm}
%Distinctive of our approach with respect to past studies
%are the normal fluid forcing scheme illustrated in Section~, the numerical distribution 
%of the normal fluid -- superfluid vortices mutual interaction force on computational grid points,
%described in Section and the parallel implementation of the code.

%%%%%%% EQUATIONS OF MOTIONS  %%%%%%%

\section{Equations of Motions}
\label{sec:eqs}

\subsection{Helium~II}

If the temperature of liquid helium is lowered
below $T_\lambda=2.17~\rm K$ at saturated vapour pressure, 
a second order phase transition takes place to another
liquid phase, known as helium~II, which exhibits
quantum-mechanical effects (unlike the high-temperature phase above
$T_{\lambda}$, called
helium~I, which can be adequately described as an ordinary Newtonian fluid).
In the absence of vortex lines, the motion of helium~II is well
described by the \textit{two-fluid model} of Landau and Tisza.
This model describes helium~II 
as the intimate mixture of two co-penetrating fluid
components \cite{tisza-1938,landau-1941,london-1954,donnelly-1991,
nemirovskii-2013}, which can be accelerated by temperature and/or pressure
gradients: 
the viscous \textit{normal fluid} and the inviscid \textit{superfluid}. Each
component is characterised by its own kinematic and thermodynamic fluid 
variables. In this description, the total density of helium~II,
$\rho$, is the sum of the partial densities $\rho_n$ and $\rho_s$ of the
normal fluid and the superfluid:
$\rho=\rho_n + \rho_s$ 
(hereafter we use the subscripts '$n$' and '$s$' to refer to
normal fluid and superfluid components, respectively). 
While $\rho$ is approximately independent of temperature 
for $T < T_{\lambda}$, the superfluid and normal fluid densities 
are strongly temperature dependent: 
in the limit $T \to T_\lambda$, helium~II becomes
entirely normal ($\rho_n/\rho \to 1$), whereas in the limit
$T \to 0$ it becomes entirely superfluid ($\rho_s/\rho \to 1$). In practice,
as the normal fluid density decreases quite rapidly with decreasing
temperature,
a helium sample at $T \leq 1~\rm K$ is effectively a pure
superfluid ($\rho_s/\rho \geq 0.99$).

Loosely speaking, the superfluid component corresponds
to the quantum ground state of the system governed by a
macroscopic complex wavefunction $\Psi$,
and the normal fluid corresponds to thermal excitations
(strictly speaking, this identification of the
superfluid component with the condensate
is not correct: helium~II is a liquid of interacting bosons, not a weakly 
interacting gas). At temperatures of the most experimental interest,
the mean free path of the thermal excitations is short enough that the
normal fluid behaves like an ordinary (classical) 
viscous fluid with non-vanishing dynamic viscosity and 
entropy per unit mass, $\eta$ and 
$s$ respectively. In contrast, the superfluid component 
is inviscid and incapable of carrying entropy (hence heat). 
%$\eta_s = s_s = 0$. 
Physically, the existence of two distinct velocity fields 
$\mathbf{v}_n$ and $\mathbf{v}_s$ signifies that, locally,
two simultaneous distinct movements are possible, 
% without, however, being able 
%to distinguish the velocity field obeyed by the single particles. 
%Helium-II \red{is therefore} a quantum fluid capable of executing 
%\textit{two motions at once}, each
%of which involves its own effective mass, 
even though individual helium atoms cannot
be separated into two components.

However, if helium~II rotates, or if the relative speed
$v_{ns}=\vert \mathbf{v}_n-\mathbf{v}_s \vert$ between normal fluid and
superfluid 
exceeds a critical value, \textit{superfluid vortex lines} 
nucleate, coupling the two fluids \cite{donnelly-1991,nemirovskii-2013} in
a way that invalidates the model of Landau and Tisza. 
It is instructive to
write the superfluid wavefunction $\Psi$ in terms of its amplitude
and phase, $\Psi=\vert \Psi \vert e^{i \phi}$, and use the quantum
mechanical prescription relating the velocity to the gradient of the
phase: $\mathbf{v}_s=(\hbar/m) \nabla \phi$. One finds \cite{primer2016}
that a vortex line 
is a topological defect: a one-dimensional region in three-dimensional
space where the phase $\phi$ is undefined, and the magnitude 
$\vert \Psi \vert$ vanishes exactly. Vortex lines are thus nodal lines of
the wavfunction.
The single valuedness of the wavefunction implies that the circulation 
integral $\Gamma$ 
following a closed path $C$ is either zero (if $C$ does not encircle
the vortex line), or takes the fixed value

\begin{equation}
\displaystyle
\Gamma = \oint_C \mathbf{v}_s \cdot d\mathbf{l} = \kappa,
\label{eq:circ_quant}
\end{equation}

\noindent
if $C$ encircles the vortex line, where $\kappa=h/m$ is the
\textit{quantum of circulation}. A multiply-charged vortex line carrying more
than one quantum of circulation is usually unstable, 
breaking into many singly-charged
vortex lines. Eq.~(\ref{eq:circ_quant}) also implies that 
the superfluid velocity around a straight vortex line has the 
form $v_s=\kappa/(2 \pi r)$, where $r$ is the distance to the vortex axis. 
Since the superfluid component has zero viscosity,
this azimuthal velocity field persists forever.  
Notice that the corresponding superfluid vorticity is a Dirac delta 
function defined on the vortex axis.

Physically, we can think of a vortex line as a thin tubular hole 
(centred on the vortex axis) around which
the circulation has the fixed value $\kappa$. The radius of the hole
is also fixed by quantum mechanics, and has the value 
$a_0 \approx 10^{-10}\rm m$,
which is the characteristic distance over which the amplitude of $\Psi$ drops
from its bulk value (at infinity, away from the vortex) to zero 
(on the vortex axis).

Vortex lines act as 
scattering centers for the thermal excitations (phonons and rotons) 
constituting the normal fluid \cite{hall-vinen-1956a,hall-vinen-1956b}.
This interaction produces an exchange of momentum, hence a mutual friction
force $\mathbf{F}_{ns}$, between the superfluid component and the normal 
fluid component. 
%The presence of this \textit{mutual friction} force $\mathbf{F}_{ns}$ 
%in the equations of motion of both fluid components, 
%the two fluid nature and the quantisation of circulation
%are all signatures of the quantum nature of Helium II. 
The quantisation of the superfluid circulation and the friction
force between superfluid and normal fluid make helium~II a  more
complex, richer system than the original two-fluid scenario of Landau and Tisza.

This aim of this work is to develop a suitable numerical framework to address 
simultaneously the coupled 
temporal evolution of \textit{(i)} the normal fluid velocity 
field and \textit{(ii)} the superfluid vortex lines, taking the 
friction into account.
In the following Sections \ref{subsec:normal_fluid} - \ref{subsec:vortex_motion}
we outline the equations of motions of the fluid components, the superfluid vortex 
lines and the expression of the friction force
employed in our numerical algorithm.

\subsection{The normal fluid}
\label{subsec:normal_fluid}
The normal fluid is a gas of elementary excitations 
called phonons and rotons. In the range of temperatures which
we are interested in, namely ${1.5 \rm~K} < T < T_\lambda$, 
the normal fluid density is mostly due to rotons \cite{hall-vinen-1956b}.
%whose mean free path is of the order of $10^{-8}\rm m$, 
The roton mean free path is
\cite{bdv1982} $\lambda_{mfp}=3 \eta /(\rho_n v_G)$, where
$v_G=\sqrt{2 k_B T/(\pi \mu)}$ is the roton group velocity,
$\mu=0.16 m_4$ is the roton effective mass \cite{brooks-donnelly-1977} and 
$m_4=6.65 \times 10^{-27}\rm kg$ is the helium mass; therefore
$\lambda_{mfp}$ varies from $13 \times 10^{-10}\rm m$
at $T=1.5 \rm K$ to $2.2 \times 10^{-10}\rm m$ at $T=2.15~\rm K$.
These values are much smaller than the typical intervortex
distance, $\ell \approx 10^{-6}\rm m$ to $10^{-4}\rm m$, at the flow's
small scales,
and the current experimental facilities at the flow's large scales,
which typically range from $10^{-3} \rm~ m$ to
$10^0 \rm~ m$ \cite{svancara-lamantia-2019,mastracci-etal-2019}.
The normal component can hence be effectively described as a fluid
with its own velocity $\mathbf{v}_n$, density $\rho_n$, entropy 
per unit mass $s$ and dynamic viscosity $\eta$. Given the small 
temperature gradients observed in experiments \cite{childers-tough-1976}, 
$\rho_n$, $s$ and $\eta$ can be treated as uniform and constant 
properties of the fluid. 
Furthermore, the normal fluid can be adequately treated as a 
Newtonian fluid, 
\textit{i.e.} the viscous stress tensor is linear in velocity gradients.
Stemming from these physical characteristics of the normal component, 
a long series of studies have focused on the derivation of the equations
of motion of helium~II in the presence of vortex lines, at 
length scales $\Delta$ much larger than the average inter-vortex spacing 
$\ell$ \cite{hall-vinen-1956b,bekarevich-khalatnikov-1961,
hills-roberts-1977}: these equations of motion are 
nowadays referred to as the
Hall-Vinen-Bekarevich-Khalatnikov (HVBK) equations.
In the cited circumstance where the normal fluid
undergoes  incompressible isoentropic motion with constant 
and uniform dynamic viscosity, the HVBK equations of motion for the 
normal component in presence of vortex lines coincide with the 
Navier--Stokes equations for a classical incompressible viscous 
fluid with the addition of an extra term $\mathbf{F}_{ns}$
representing the friction force:

\begin{eqnarray}
\displaystyle
\rho_n\left [ \frac{\partial \mathbf{v}_n}{\partial t} + \left ( \mathbf{v}_n \cdot \nabla \right ) \mathbf{v}_n \right ] & = & 
-\frac{\rho_n}{\rho}\nabla p -\rho_s s \nabla T + \eta \nabla^2 \mathbf{v}_n + \mathbf{F}_{ns} \label{eq:NS} \; \; , \\[3mm]
\nabla \cdot \mathbf{v}_n & = & 0 \label{eq:vn_incompress} \; \; ,
\end{eqnarray}

\noindent
where $p$ is pressure.
% and Eq. (\ref{eq:vn_incompress}) expresses the incompressibility of the normal fluid flow. 
Considering hereafter only isothermal helium~II and introducing 
characteristic units
of length and time, respectively $\lambda$ and $\tau$, Eqs.~(\ref{eq:NS}) 
and (\ref{eq:vn_incompress}) can be made non-dimensional as follows:

\begin{eqnarray}
\displaystyle
\frac{\partial \widetilde{\mathbf{v}}_n}{\partial \widetilde{t}} + \left ( \widetilde{\mathbf{v}}_n \cdot \widetilde{\nabla} \right ) \widetilde{\mathbf{v}}_n & = & 
-\widetilde{\nabla}\widetilde{\left (\frac{p}{\rho} \right )} + \frac{\nu_n}{(\lambda^2/\tau)} \widetilde{\nabla}^2 \widetilde{\mathbf{v}}_n + 
\left (\frac{\mathbf{F}_{ns}}{\rho_n}\right )\frac{\tau^2}{\lambda} 
	\label{eq:NS_nondim} \; \; , \\[3mm]
\widetilde{\nabla} \cdot \widetilde{\mathbf{v}}_n & = & 0 \label{eq:vn_incompress_nondim} \; \; ,
\end{eqnarray}

\noindent
where $\widetilde{\cdot}$ indicates non-dimensional quantities 
and $\nu_n=\eta/\rho_n$ is the kinematic viscosity of the normal fluid. 
The numerical schemes for the integration of Eqs.~\refeq{eq:NS_nondim} 
and \refeq{eq:vn_incompress_nondim} and 
%the discussion concerning the open questions on 
the numerical handling 
of the friction force $\mathbf{F}_{ns}$ are 
%illustrated 
discussed in Sections~\ref{subsec:NS} and \ref{subsec:extrap},
respectively. 

\subsection{The superfluid}
\label{subsec:superfluid}

The complete set of HVBK equations consists of Eqs.~(\ref{eq:NS}) 
and (\ref{eq:vn_incompress}) supplemented with the equations of motion 
of the superfluid component which, in the incompressible approximation
and neglecting the tension force, are as follows:

\begin{eqnarray}
\displaystyle
\rho_s\left [ \frac{\partial \mathbf{v}_s}{\partial t} + \left ( \mathbf{v}_s \cdot \nabla \right ) \mathbf{v}_s \right ] & = & 
-\frac{\rho_s}{\rho}\nabla p + \rho_s s \nabla T - \mathbf{F}_{ns} \label{eq:superfluid} \; \; , \\[3mm]
\nabla \cdot \mathbf{v}_s & = & 0 \label{eq:vs_incompress} \; \; .
\end{eqnarray}

\noindent
As already said,
the superfluid vorticity $\bm{\omega}_s=\nabla \times \mathbf{v}_s$ is 
confined to the vortex-lines which can effectively be described
as one-dimensional objects because the vortex
core size $a_0 \approx 10^{-10}\rm~m$ 
is several orders of magnitude smaller than the flow length scales 
probed in the present work ($\approx 10^{-5} \rm~m$ to
$10^{-6} \rm~m$). The vortex lines can hence be treated as 
parametrised space curves 
$\mathbf{s}(\xi,t)$ in a three-dimensional domain, where $\xi$ is 
arc-length and $t$ is time. Within this mathematical description 
of vortex lines,
the superfluid vorticity $\bm{\omega}_s$ can be expressed in terms of 
$\delta$-distributions, as follows

\begin{equation}
\displaystyle
\bm{\omega}_s (\mathbf{x},t)= \kappa \oint_\mathcal{L} \mathbf{s}'(\xi,t) \delta (\mathbf{x} - \mathbf{s}(\xi,t)) d\xi \;\;\;\; , \label{eq:omega_distr}
\end{equation}

where 
%$\displaystyle \mathbf{s}'(\xi,t) =\frac{\partial \mathbf{s}(\xi,t)}{\partial \xi}$ 

\begin{equation}
\displaystyle 
\mathbf{s}'(\xi,t) =\frac{\partial \mathbf{s}(\xi,t)}{\partial \xi},
\end{equation}

\noindent
is the unit tangent vector to the curve $\mathbf{s}(\xi,t)$ and
$\mathcal{L}$ indicates the whole vortex configuration.

The superfluid velocity may hence be expressed in terms of the spatial distribution of vortex lines $\mathbf{s}(\xi,t)$ via the Biot-Savart
integral \cite{schwarz-1978}, \textit{i.e.}
\begin{equation}
\displaystyle
\mathbf{v}_s (\mathbf{x},t)= \nabla \phi (\mathbf{x},t) + 
\frac{\kappa}{4\pi } \oint_\mathcal{L}\frac{\mathbf{s}'(\xi,t)\times (\mathbf{x}-\mathbf{s}(\xi,t))}{|\mathbf{x}-\mathbf{s}(\xi,t)|^3} d\xi \;\;\;\; ,\label{eq:vs}
\end{equation}
where $\nabla \phi (\mathbf{x},t)$ is the potential flow arising from the macroscopic boundary conditions. The corresponding non-dimensional
equation is straightforwardly obtained and reads as follows:
\begin{equation}
\displaystyle
\widetilde{\mathbf{v}}_s (\widetilde{\mathbf{x}},\tilde{t})= \widetilde{\nabla} \widetilde{\phi} (\widetilde{\mathbf{x}},\tilde{t}) + 
\frac{\kappa}{4\pi (\lambda^2/\tau)} \oint_\mathcal{L}
\frac{\mathbf{s}'(\widetilde{\xi},\tilde{t})\times (\widetilde{\mathbf{x}}-\widetilde{\mathbf{s}}(\widetilde{\xi},\tilde{t}))}{|\widetilde{\mathbf{x}}-\widetilde{\mathbf{s}}(\widetilde{\xi},\tilde{t})|^3} d\widetilde{\xi} \;\;\;\; . \label{eq:vs_no_dim}
\end{equation}
\noindent
The numerical computation of Eq. \refeq{eq:vs_no_dim} is addressed in Section \ref{subsec:VFM}.

\subsection{The  friction force}
\label{subsec:mutual_friction}

Historically, three distinct theoretical frameworks 
have been employed in the literature to model  
the interaction between vortex lines 
and normal fluid: (i) the \textit{coarse-grained} philosophy,  
(ii) the \textit{local} approach, and (iii) the
\textit{fully-coupled local} approach. The first approach probes the
flow at length scales $\Delta$ much larger than the average 
inter-vortex spacing $\ell$; it was originally derived in the 
pioneering studies of Hall and Vinen \cite{hall-vinen-1956a,hall-vinen-1956b} 
and successively led to the HVBK theoretical formulation 
\cite{bekarevich-khalatnikov-1961,hills-roberts-1977}. 
The second approach addresses helium~II dynamics at the scale of 
individual vortex line elements of length $\delta < \ell$; it was first
employed by Schwarz \cite{schwarz-1978,schwarz-1985,schwarz-1988} 
and later reformulated by Kivotides, Barenghi and Samuels
\cite{kivotides-barenghi-samuels-2000}.
%,idowu-willis-barenghi-samuels-2000,idowu-kivotides-barenghi-samuels-2000}.
The third approach is a modification of the second that includes
the effects of the local back-reaction of the vortex lines on the normal 
fluid.

%\subsubsection{Coarse-grained mutual friction}
\begin{enumerate}[label=(\roman*)]
\item{Coarse-grained friction}

At length scales $\Delta \gg \ell$, the discrete, singular nature of the
superfluid vorticity field is lost. As a result, the averaged superfluid 
velocity and vorticity
fields, indicated hereafter by $\avg{\mathbf{v}_s}$ 
and $\avg{\bm{\omega_s}}$ respectively, are 
\textit{continuous} fields, smoothly varying on the macro-scale $\Delta$. 
At these scales, Hall and Vinen deduced the following
expression for the  
friction force $\avg{\mathbf{F}_{ns}}$ acting per unit volume on 
the normal fluid in a bucket of rotating helium~II:

\begin{equation} 
\displaystyle
\avg{\mathbf{F}_{ns}} = B\frac{\rho_n \rho_s}{2\rho\avg{\omega_s}}\avg{\bm{\omega_s}} \times 
\left [\avg{\bm{\omega_s}} \times \left ( \avg{\mathbf{v_n}} - \avg{\mathbf{v_s}}\right ) \right ] +
B'\frac{\rho_n \rho_s}{2\rho}\avg{\bm{\omega_s}} \times \left ( \avg{\mathbf{v_n}} - \avg{\mathbf{v_s}}\right )\;\;\; , \label{eq:F_ns_HV}
\end{equation}

\noindent
where $\avg{\mathbf{v_n}}$ is the coarse-grained normal fluid velocity 
and $B$ and $B'$ are friction coefficients determined
experimentally at scales $\Delta$ via second sound attenuation 
measurements in a rotating cryostat 
\cite{hall-vinen-1956a,hall-vinen-1956b}. 
Eq.~(\ref{eq:F_ns_HV}) was generalised 
for non-straight vortex configurations by Bekarevich 
and Khalatnikov \cite{bekarevich-khalatnikov-1961}.
In this form (which contains a vortex tension term), the HVBK equations
were applied to Couette flow
\cite{BarenghiJones1988,Barenghi1992,Henderson1995}, predicting
with success the transition from Couette flow to Taylor vortex flow 
and the weakly nonlinear regime at higher velocities beyond the transition. 
Being laminar, these flows satisfy the assumption behind
the derivation of the HVBK equations that the vortex lines are locally
polarised. The application of the
HVBK equations to turbulent flows is less straighforward, as
the vortex lines are polarised only partially (most vortex lines
are at random directions with respect to each other), hence
the relation $\vert \avg{\bm{\omega_s}} \vert = \kappa L$ between the 
vortex line density $L$ and the coarse-grained superfluid vorticity is
only approximate. Nevertheless, the HVBK equations have been used
to model the turbulent cascade in helium~II \cite{Roche2009}.

%\subsubsection{Local utual friction}
\item{Local friction}

%In order to make progress in the understanding of Helium II 
%turbulence and, more in particular, i
Developing the vortex filament method,
Schwarz derived an expression for the force
per unit length $\avg{\mathbf{f}_{sn}}$ exerted by the normal fluid 
on a \textit{single} vortex line element of length $\delta < \ell$, 
position $\mathbf{s}(\xi,t)$ and unit tangent vector 
$\mathbf{s}'$ \cite{schwarz-1978}:

\begin{equation} 
\displaystyle
\avg{\mathbf{f}_{sn}} = - \alpha \rho_s \kappa \; \mathbf{s}' \times \left [  \mathbf{s}' \times \left ( \widehat{\mathbf{V}}_n - \mathbf{v}_s\right ) \right ] -
\alpha'\rho_s \kappa \; \mathbf{s}' \times \left ( \widehat{\mathbf{V}}_n - \mathbf{v}_s\right )
\;\;\; , \label{eq:F_ns_Schwarz}
\end{equation} 

\noindent
where $\alpha = B\rho_n/(2\rho)$, $\alpha' = B'\rho_n/(2\rho)$, 
%($B$ and $B'$ being the mutual friction coefficients introduced 
%in Eq. (\ref{eq:F_ns_HV})),
$\mathbf{v}_s = \mathbf{v}_s (\mathbf{s}(\xi,t),t)$ 
defined by the Biot - Savart integral in Eq. (\ref{eq:vs}) and 
$\widehat{\mathbf{V}}_n = \widehat{\mathbf{V}}_n (\mathbf{s}(\xi,t),t)$ 
is a \textit{prescribed} normal fluid flow. 

Eq. (\ref{eq:F_ns_Schwarz}) is based on the following 
two important assumptions. 
Firstly, it neglects the back reaction of the superfluid vortex motion
on the flow of the normal fluid. It assumes in fact, that each vortex line 
element feels the normal fluid flow $\avg{\mathbf{v_n}}$, \textit{i.e.} a 
macroscopic velocity field averaged over a region containing 
many vortex lines ($\Delta \gg \ell$). The field $\avg{\mathbf{v_n}}$ 
is thus determined by the macroscopic boundary conditions of the flow 
which is investigated and is entirely decoupled from the evolution
of the superfluid vortex tangle. 
As a consequence, in the framework elaborated by Schwarz, 
the normal fluid velocity field may be prescribed \textit{a priori}, 
depending on the
fluid dynamic characteristics of the system studied: the explicit 
presence of the prescribed flow $\widehat{\mathbf{V}}_n$ 
in Eq.~(\ref{eq:F_ns_Schwarz})
underlies this aspect. 
The second assumption, strongly linked to the previous, is the use of 
macroscopic coefficients $\alpha$ and $\alpha'$ for the calculation of 
the friction acting on a \textit{single} vortex element:
$\alpha$ and $\alpha'$ are in fact simply redefinitions
of the  friction coefficients $B$ and $B'$ determined at 
large scales $\Delta$ in  
a very particular vortex line configuration (a lattice of straight 
vortex lines in a rotating bucket)
\cite{hall-vinen-1956a,hall-vinen-1956b}. 
Eq.~(\ref{eq:F_ns_Schwarz}), on the contrary, is intended to 
describe the force experienced by a vortex line element in a 
turbulent tangle. The measured values of $\alpha$ and $\alpha'$ are displayed in Fig.\ref{fig:PhysParam}. 

This decoupling of the normal fluid flow from the vortex lines motion 
(and, hence, the possibility of imposing arbitrarily 
a priori the normal fluid velocity field 
$\widehat{\mathbf{V}}_n(\mathbf{x},t)$ felt by the vortices)
confers to the theoretical framework pioneered by Schwarz 
a \textit{kinematic} character: the evolution of the vortex tangle 
is determined 
for a given imposed normal flow. This local kinematic approach
has been extensively employed in past studies to shed light
on fundamental aspects of superfluid turbulence. In particular, 
various models of imposed normal flow $\widehat{\mathbf{V}}_n$ 
have been studied: uniform
\cite{schwarz-1988,adachi-fujiyama-tsubota-2010,baggaley-sherwin-barenghi-sergeev-2012,sherwin-barenghi-baggaley-2015},
parabolic \cite{aarts-dewaele-1994,baggaley-laizet-2013,khomenko-etal-2015,baggaley-laurie-2015},
Hagen--Poiseuille and tail--flattened flows \cite{yui-tsubota-2015},
vortex tubes \cite{samuels-1993}, 
ABC flows \cite{barenghi-samuels-bauer-donnelly-1997},
frozen normal fluid vortex tangles \cite{kivotides-2006}, 
random waves \cite{sherwin-barenghi-baggaley-2015},
time--frozen snapshots of turbulent solutions of 
Navier--Stokes equations
\cite{baggaley-laizet-2013,sherwin-barenghi-baggaley-2015,baggaley-laurie-2015} and 
time--dependent homogeneous and isotropic turbulent solutions of linearly forced
Navier--Stokes equations \cite{morris-koplik-rouson-2008}.

\item{Self-consistent local friction}
 
The self-consistent local approach was formulated in order to take into account 
the back reaction of the vortex lines onto the normal fluid 
\cite{kivotides-barenghi-samuels-2000,idowu-willis-barenghi-samuels-2000,
idowu-kivotides-barenghi-samuels-2000},  by modelling
the dragging of the roton gas of excitations constituting 
the normal fluid by the vortex lines moving relative to it.
Here the velocity field varies 
at length scales smaller than the average inter-vortex spacing $\ell$ 
and depends on the evolution of the vortex tangle: it cannot be prescribed, 
but has to be determined via the integration of the incompressible 
Navier--Stokes equations \refeq{eq:NS} and \refeq{eq:vn_incompress}. 
This mismatch between the local normal fluid velocity $\mathbf{v}_n$ 
perturbed by superfluid vortices 
%(which we will hereafter 
%refer to as \textit{mesoscopic} normal velocity) 
and the macroscopic flow $\widehat{\mathbf{V}}_n$ implies the 
necessity of re-determining the friction coefficients at these 
scales, now related to the
local cross sections between the roton gas and the vortex lines
\cite{idowu-kivotides-barenghi-samuels-2000}. 

This self-consistent local approach was first employed to 
investigate the normal fluid velocity 
field induced by simple vortex configurations, namely
vortex rings \cite{kivotides-barenghi-samuels-2000} and 
vortex lines \cite{idowu-willis-barenghi-samuels-2000}; 
later, it was also used to study both 
the forcing of the normal fluid flow by decaying superfluid vortex tangles 
\cite{kivotides-2005,kivotides-2007} and, vice-versa, the stretching 
of an initially small superfluid vortex tangles by either
a decaying turbulent normal fluid 
\cite{kivotides-2011,kivotides-2014,kivotides-2015jltp} 
or decaying normal fluid structures such as as rings 
\cite{kivotides-2015epl}
and Hopf links \cite{kivotides-2018}.

\end{enumerate}

The present work employs the most recent formulation 
of the third approach \cite{kivotides-2018}, 
slightly revisited.  We model each vortex line element 
as a cylinder of radius $a_0$ and 
length $\delta$ (coinciding with the spatial discretisation 
of vortex lines, see Section \ref{subsec:VFM} for details), 
where $\delta \gg a_0$. The 
very small vortex core size $a_0$ implies that when a superfluid vortex 
is in relative motion with respect to the normal fluid, it generates 
a low Reynolds number normal flow around itself. 
Typical values of the Reynolds number associated to 
this vortex-generated normal flow
in helium experiments are $10^{-5}$ to $10^{-4}$.
From the theory of low Reynolds number flows,
\cite{kaplun-lagerstrom-1957,kaplun-1957,proudman-pearson-1957,vandyke-1975}, 
the force per unit length $\mathbf{f}_D$ which the normal fluid exerts on the 
vortex line is

\begin{equation}
\displaystyle
\mathbf{f}_D = - D\,\mathbf{s}' \times \left [ \mathbf{s}' \times \left ( \mathbf{v}_n - \dot{\mathbf{s}} \right ) \right ] \;\; ,
\label{eq:f_D_Kivotides}
\end{equation}

\noindent
where the coefficient $D$ (not to be confused with the friction
coefficient discussed in Ref.~\cite{bdv1982}) is

\begin{equation}
\displaystyle D= \frac{4\pi\rho_n\nu_n}{[\frac{1}{2}-\gamma-\ln\left 
( \frac{|\mathbf{v}_{n_{\perp}} - \dot{\mathbf{s}}|a_0}{4\nu_n}\right )]},
\end{equation}

\noindent 
and $\gamma=0.5772$ is the Euler-Mascheroni constant.
% and $\mathbf{v}_{n_{\perp}}$ is the normal fluid velocity perpendicular to $\mathbf{s}'$.

\noindent 
Furthermore, 
$\displaystyle \dot{\mathbf{s}} = \partial \mathbf{s}(\xi,t)/\partial t$ 
is the velocity of the vortex line 
(see next paragraph \ref{subsec:vortex_motion}),  
$\mathbf{v}_n$ is evaluated on the vortex line element, that is to say
$\mathbf{v}_n=\mathbf{v}_n(\mathbf{s}(\xi,t),t)$ using
the interpolation schemes described and tested in 
Section~\ref{subsec:interp}, and the quantity
$\mathbf{v}_{n_{\perp}}$
indicates the component of the normal fluid velocity 
lying on a plane orthogonal to $\mathbf{s}'$. 

The expression \refeq{eq:f_D_Kivotides} for the viscous force 
$\mathbf{f}_D$ acting on the vortex line differs from the most recent approach \cite{kivotides-2018} as it is independent of the discretisation  $\delta$ on the lines.
Including further the Iordanskii force $\mathbf{f}_I~=~-~\rho_n\kappa\; \mathbf{s}' \times \left ( \mathbf{v}_n - \dot{\mathbf{s}} \right )$ \cite{volovik-1995,thompson-stamp-2012}, 
%in the theoretical framework we employ in the present study, 
the total force per unit length $\mathbf{f}_{sn}$ acting on the superfluid vortices stemming from the interaction with the normal fluid is as follows
\begin{equation}
\displaystyle
\mathbf{f}_{sn} (\mathbf{s})= \mathbf{f}_D + \mathbf{f}_I = - D \,\mathbf{s}' \times \left [ \mathbf{s}' \times \left ( \mathbf{v}_n - \dot{\mathbf{s}} \right ) \right ]
- \rho_n\kappa\; \mathbf{s}' \times \left ( \mathbf{v}_n - \dot{\mathbf{s}} \right ) \;\; .
\label{eq:f_ns_Kivotides}
\end{equation}

\noindent
In conclusion of this section, for the sake of completeness it is important to mention that also Eq.~\refeq{eq:F_ns_Schwarz} has been employed to take into account 
the back reaction of the superfluid vortices on the normal fluid, by averaging $\avg{\mathbf{f}_{sn}}$ on 
a normal fluid grid cell containing many vortex line elements \cite{galantucci-sciacca-barenghi-2015,galantucci-sciacca-barenghi-2017,yui-etal-2018}.

\subsection{The motion of vortex lines}
\label{subsec:vortex_motion}

The derivation of the equations of motion of the vortex lines is
straightforward once Eq. (\ref{eq:f_ns_Kivotides}) is taken into account. 
Since the vortex core is much smaller then any
other scales of the flow, the vortex inertia can be neglected. 
As a consequence, the sum of all
the forces acting on a vortex vanishes. 
Since the vortex line is effectively like a small cylinder in an inviscid
fluid (the superfluid) surrounded by a circulation and in relative
motion with respect to a background flow, it suffers a Magnus force
per unit length which is

\begin{equation}
\displaystyle
\mathbf{f}_{M} = \rho_s \kappa \; \mathbf{s}' \times \left ( \dot{\mathbf{s}} - \mathbf{v}_s \right ) \; \; ,
\label{eq:f_M}
\end{equation}

\noindent
where $\mathbf{v}_s$ is evaluated on the vortex, 
$\mathbf{v}_s=\mathbf{v}_s(\mathbf{s}(\xi,t),t)$.
Setting the sum of all forces acting on the vortex line equal to zero, 
$\mathbf{f}_{sn} + \mathbf{f}_{M} = 0$, and employing 
Eqs.~\refeq{eq:f_ns_Kivotides} and \refeq{eq:f_M}, we have

\begin{equation}
\displaystyle
\mathbf{s}' \times \left [ -D \, \mathbf{s}' \times \left ( \mathbf{v}_n - \dot{\mathbf{s}} \right ) - \rho_n\kappa\; \left ( \mathbf{v}_n - \dot{\mathbf{s}} \right ) +
\rho_s \kappa \; \left ( \dot{\mathbf{s}} - \mathbf{v}_s \right )\right ] =  0 \; \; .
\label{eq:vortex_motion_initial}
\end{equation}

\noindent
Assuming that each vortex line element moves orthogonally to its unit tangent vector, \textit{i.e.} $\dot{\mathbf{s}} \cdot \mathbf{s}' = 0$, Eq. \refeq{eq:vortex_motion_initial} leads 
\cite{idowu-kivotides-barenghi-samuels-2000,kivotides-2018}  to the following
equation of motion for $\dot{\mathbf{s}}(\xi,t)$ 

\begin{eqnarray}
\displaystyle
\dot{\mathbf{s}} & = & - \mathbf{s}' \times \left \{ \mathbf{s}' \times \left [ \left ( 1 + \alpha \right ) \mathbf{v}_s - \alpha  \mathbf{v}_n \right ] \right \} 
+ \beta \mathbf{s}' \times  \left ( \mathbf{v}_n - \mathbf{v}_s \right )     \nonumber \\[2mm]
& = &  \mathbf{v}_{s_{\perp}}
+ \beta \mathbf{s}' \times  \left ( \mathbf{v}_n - \mathbf{v}_s \right )
+\beta' \mathbf{s}' \times \left [ \mathbf{s}' \times \left ( \mathbf{v}_n - \mathbf{v}_s \right ) \right ] \; \; ,
\label{eq:vortex_motion_final}
\end{eqnarray}

\noindent
where 
\begin{equation}
\displaystyle
\beta =  \frac{a}{(1+b)^2 + a^2} > 0,
\end{equation}

\begin{equation}
\displaystyle 
\beta' = - \frac{b(1+b) + a^2}{(1+b)^2 + a^2} < 0,
\end{equation}

\begin{equation}
\displaystyle 
a=\frac{D}{\rho_s \kappa}= 4\pi\;
\left(\frac{\rho_n} {\rho_s}\right) 
\left( \frac{\nu_n}{\kappa} \right) \;
\frac{1}{[\frac{1}{2}-\gamma-\ln\left ( \frac{|\mathbf{v}_{n_{\perp}} - \dot{\mathbf{s}}|a_0}{4\nu_n}\right )]}\;\;
\end{equation}

\noindent
and 

\begin{equation}
\displaystyle 
b=\frac{\rho_n}{\rho_s}.
\end{equation}

\noindent
From the physical point of view, the motion of the vortices 
is hence governed only by temperature,
which determines $\rho_n/\rho_s$ and $\nu_n/\kappa$.

\subsection{The self-consistent model}

Our model of fully self-consistent motion of helium~II
at finite temperature comes from gathering the dimensionless equations
\eqref{eq:NS_nondim}, \eqref{eq:vn_incompress_nondim}, 
\eqref{eq:vs_no_dim}, \eqref{eq:f_ns_Kivotides} and 
\eqref{eq:vortex_motion_final}. The dimensional scaling factors 
$\tau$ and $\lambda$ can be chosen such that the integral 
time and length scales of the normal fluid are of order one. 
In particular we set
\begin{equation}
\nu_n^0=\frac{\nu_n \tau}{\lambda^2}, \qquad
\Gamma=\frac{\kappa}{\nu_n}. 
\label{Eq:nuNume}
\end{equation}

The dimensionless viscosity $\nu_n^0$ is set to properly resolve 
the small scales of the normal fluid. Note that $\Gamma$ is dimensionless 
and depends on temperature. When comparing the
simulations to  experiments, physical time and length scales are 
recovered from Eq.\eqref{Eq:nuNume} by choosing the unit of length 
of the system, $\lambda$.
With this notation, our self-consistent model, written 
in dimensionless form (after dropping tildes), reads:

\begin{eqnarray}
\displaystyle
\label{Eq:SelfConstModelEqVn}
&& \frac{\partial \mathbf{v}_n}{\partial t} + \left ( \mathbf{v}_n \cdot \nabla \right ) \mathbf{v}_n  =  
-\nabla \left (\frac{p}{\rho_n} \right ) + \nu_n^0 \nabla^2 \mathbf{v}_n +\frac{1}{\rho_n}\oint_\mathcal{L} \delta(\mathbf{x-s}) \mathbf{f}_{ns}(\mathbf{s}) \mathrm{d}\xi+\frac{1}{\rho_n}\mathbf{F}_{\rm ext}\,\\
&& \nabla \cdot \mathbf{v}_n=0 \\[2mm]
\label{Eq:SelfConstModelEqFns}
&&\frac{1}{\rho_n}\mathbf{f}_{ns} ({\bf s} )= - \frac{1}{\rho_n}\mathbf{f}_{sn} (\mathbf{s}) =- \nu_n^0\Gamma \mathbf{s}' \times \left ( \dot{\mathbf{s}} - \mathbf{v}_n \right ) \;  
- \nu_n^0 D^0 \mathbf{s}' \times \left [ \mathbf{s}' \times \left ( \dot{\mathbf{s}} - \mathbf{v}_n \right ) \right ] \\[2mm]
\label{Eq:SelfConstModelEqVFM}&&\dot{\mathbf{s}}= \mathbf{v}_{s_{\perp}}(\mathbf{s})
+ \beta \,\mathbf{s}' \times  \left ( \mathbf{v}_n(\mathbf{s}) - \mathbf{v}_s(\mathbf{s}) \right )
+\beta' \,\mathbf{s}' \times \left [ \mathbf{s}' \times \left ( \mathbf{v}_n(\mathbf{s}) - \mathbf{v}_s(\mathbf{s}) \right ) \right ]\\[2mm]
%\mathbf{s}' \times \left \{ \mathbf{s}' \times \left [ \left ( 1 + \alpha \right ) \mathbf{v}_s(\mathbf{s}) - \alpha  \mathbf{v}_n(\mathbf{s}) \right ] \right \} 
%+ \beta \mathbf{s}' \times  \left ( \mathbf{v}_n(\mathbf{s}) - \mathbf{v}_s(\mathbf{s}) \right )    \\[2mm]
%
\label{Eq:SelfConstModelEqVS}
&&\mathbf{v}_s (\mathbf{x},t)= \nabla \phi (\mathbf{x},t) + 
\frac{\Gamma \nu_n^0}{4\pi } \oint_\mathcal{L}\frac{\mathbf{s}'(\xi,t)\times (\mathbf{x}-\mathbf{s}(\xi,t))}{|\mathbf{x}-\mathbf{s}(\xi,t)|^3} d\xi
\end{eqnarray}
with 
\begin{equation}
\displaystyle 
D^0= \frac{4\pi}{[\frac{1}{2}-\gamma-\ln\left ( \frac{|\mathbf{v}_{n_{\perp}} - \dot{\mathbf{s}}|a_0}{4\nu_n}\right )]}
\end{equation}

\begin{figure}
\begin{center}
%\vspace*{5cm}       % Give the correct figure height in cm
\includegraphics[width=0.32\textwidth]{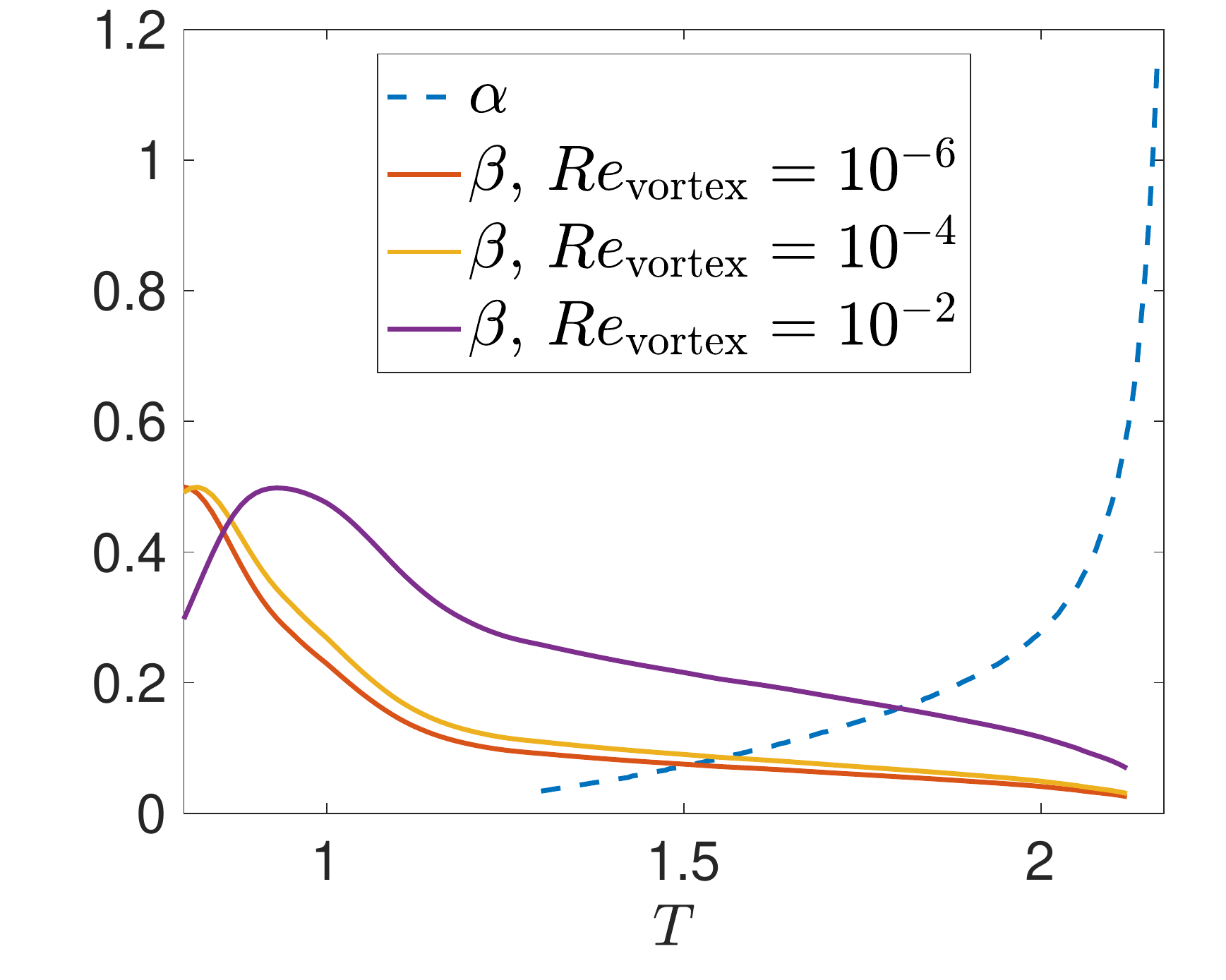} 
\includegraphics[width=0.32\textwidth]{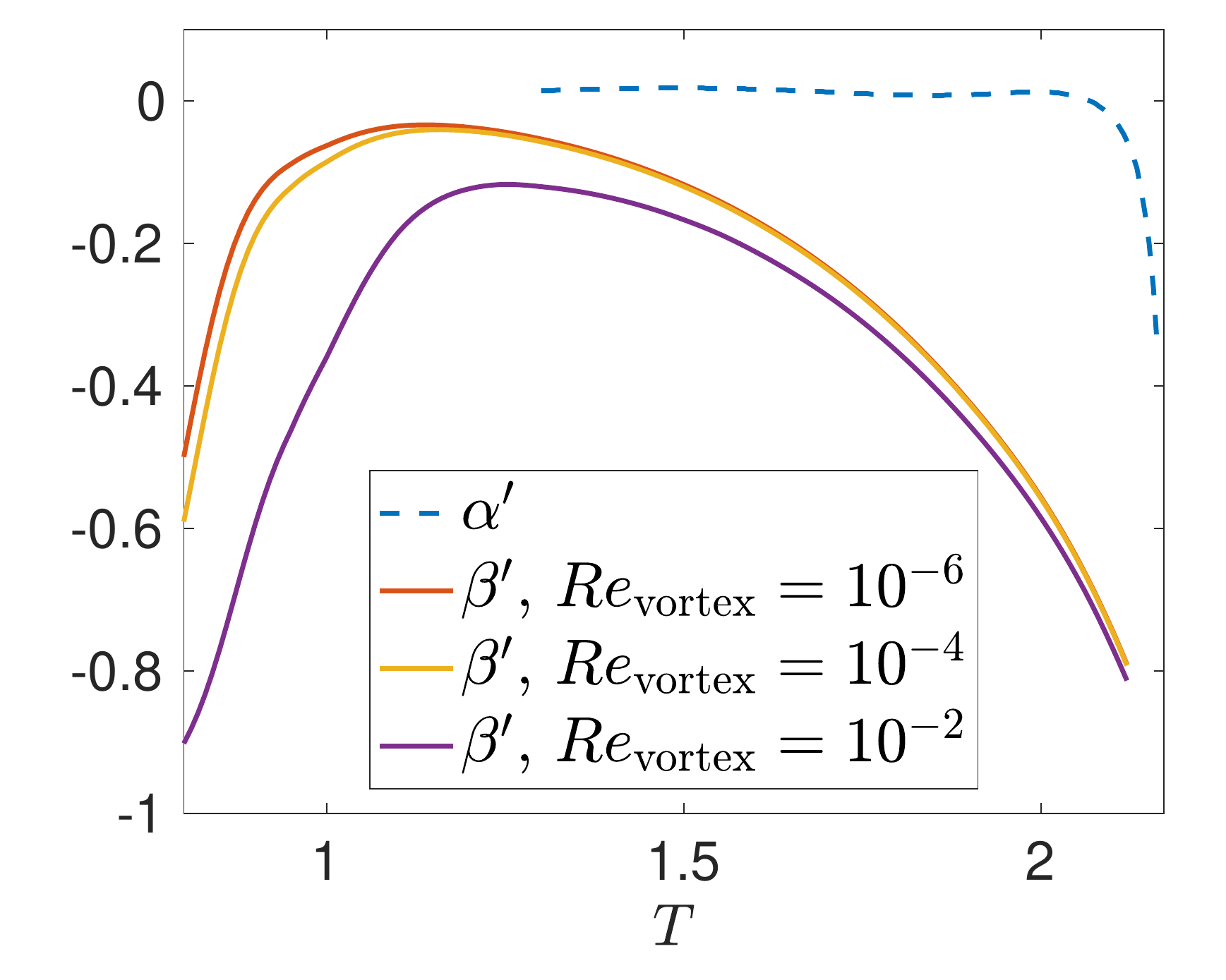} 
\includegraphics[width=0.32\textwidth]{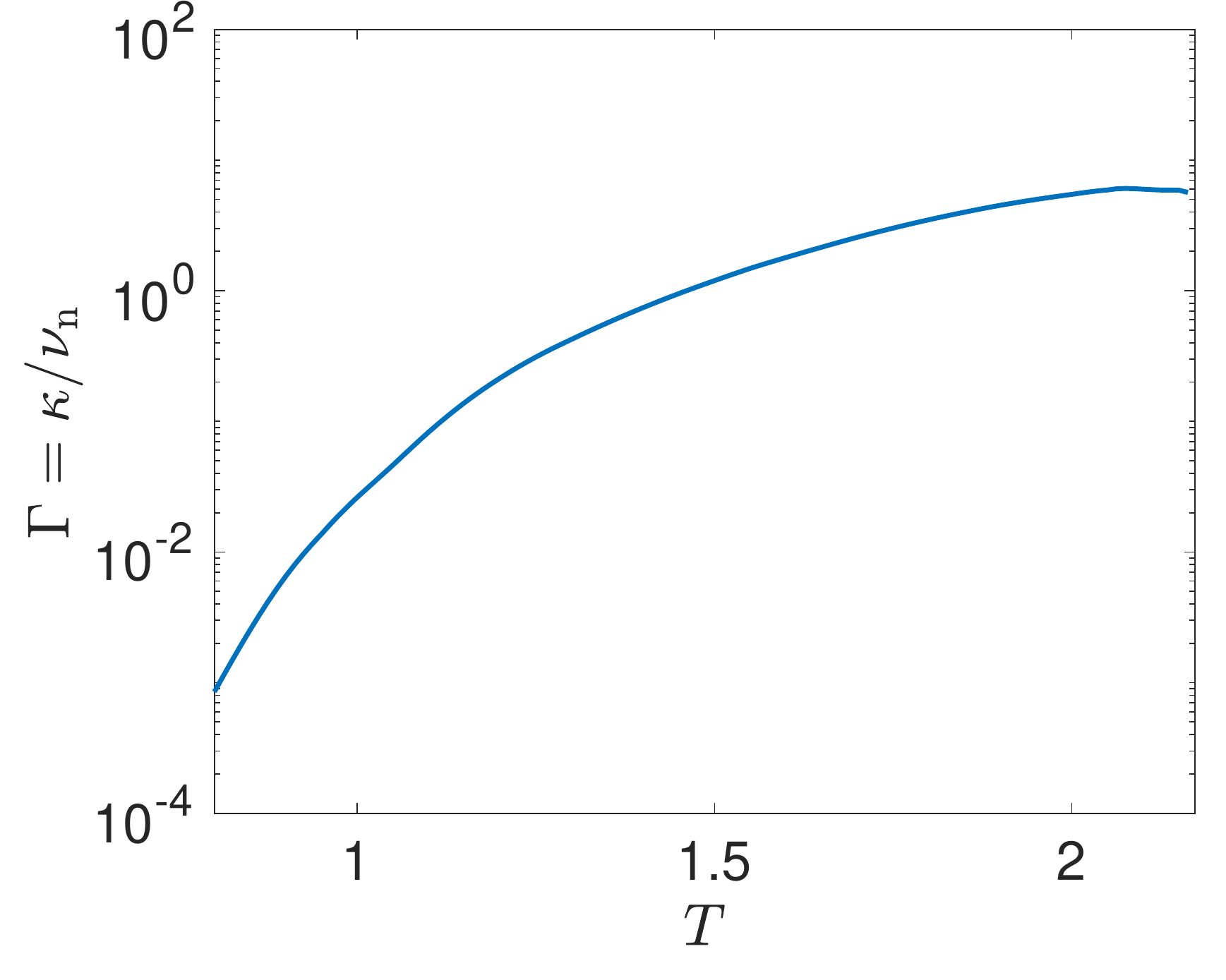} 
\caption{Dimensionless friction parameters $\alpha$ and $\beta$  (left), $\alpha'$ and $\beta'$ (centre)
and $\Gamma$ (right) as a function of temperature $T$ and,
for $\beta$ and $\beta'$, typical vortex Reynolds numbers, defined as 
$Re_{\rm vortex}={|\mathbf{v}_{n_{\perp}} - \dot{\mathbf{s}}|a_0}/{4\nu_n}$.
}
\label{fig:PhysParam}       % Give a unique label
\end{center}
\end{figure}

\noindent 
In Eq.\eqref{Eq:SelfConstModelEqVn} we also added the external force ${\mathbf{F}_{\rm ext} }$ in order to sustain turbulence in the normal fluid. 
We recall that the physical parameters $\beta,\beta'$ and $\Gamma$ depend only on temperature. 
Their behaviour is displayed on Fig.\ref{fig:PhysParam} using the tabulated values of 
$\rho_n,\rho_s$ and $\nu_n$ from reference \cite{Donnelly1998}. For the sake of completeness, in 
Fig.\ref{fig:PhysParam} we also report the temperature dependence  
of the mutual friction coefficients $\alpha$ and $\alpha'$ introduced by Schwarz \cite{schwarz-1978}
while modeling locally the mutual friction force in presence of a prescribed normal fluid flow 
(cfr. Eq.~\refeq{eq:F_ns_Schwarz}).

Finally note that for a given temperature, there are two more dimensionless parameters
\begin{eqnarray}
Re=\frac{ v_n^{\rm rms}L}{\nu_n} \quad&{\rm and}&\quad I_{\rm turb}=\frac{v_n^{\rm rms}}{|\langle{\bf v}_n - {\bf v}_s\rangle|},
\end{eqnarray}
where $v_n^{\rm rms}$ is the root-mean-square or the characteristic normal fluid velocity fluctuation and $L$ is its integral scale. 
The Reynolds number $Re$ tells how turbulent is the normal fluid and the turbulent intensity $I_{\rm turb}$ measures the strength of
thermal counterflows with respect to normal fluid turbulent fluctuations. 
Note that, alternatively, we can also use a Reynolds number 
$\displaystyle Re_\lambda=\frac{v_n^{\rm rms}\lambda_T}{\nu_n}$ based on the Taylor micro scale of the flow $\lambda_T$, 
%\textit{i.e.} $Re_\lambda=\frac{v_n^{\rm rms}\lambda_T}{\nu_n}$ , that provides a 
which provides a precise definition in terms of velocity gradients and 
fluctuations \cite{Frisch1995} (the relation between the Taylor micro scale $\lambda_T$ and the 
the turbulent kinetic energy dissipation $\epsilon$ in homogenous and isotropic turbulence is as follows
$\epsilon = 15 \nu_n (v_n^{\rm rms}/\lambda_T)^2$ leading to $Re_\lambda \sim \sqrt{15}\; Re^{1/2}$).

\section{Numerical Method}
\label{sec:numerical}

The numerical integration of the self-consistent model Eqs.~(\ref{Eq:SelfConstModelEqVn}) - (\ref{Eq:SelfConstModelEqVS}) 
demands special care. In particular, the coupling between the vortex 
filament method and the Navier--Stokes equation requires 
interpolations to determine the values of the normal fluid 
at the filament position and a redistribution of the force 
%from 
$\mathbf{f}_{ns}$ onto the mesh where the normal fluid is defined. 
As we shall see at the end of this section, a careless treatment 
of the coupling may lead to spurious numerical artefacts.

In this section we describe the numerical method used to solve 
equations (\ref{Eq:SelfConstModelEqVn}) - (\ref{Eq:SelfConstModelEqVS}) 
and provide physical and numerical justifications for the
choice of the numerical scheme used to couple superfluid and normal fluid.
We start by briefly describing the numerical scheme for the 
Vortex Filament Method and for the Navier--Stokes equations,
and then proceed to the friction coupling.

%%%%%%%%%% VFM  %%%%%%%%%%%%%%%%%%%%%

\subsection{Vortex Filament Method}
\label{subsec:VFM}

Here we briefly describe the Vortex Filament Method (VFM) to determine
the time evolution of vortex lines.
For a more in-depth overview of the VFM, we point the reader to a recent 
review article \cite{hanninen-baggaley-2014} and references within.
The underlying assumption of the VFM is that vortex lines
in the superfluid component can be considered one-dimensional 
space curves around which the 
circulation is one quantum of circulation $\kappa$. 
This is a reasonable assumption in helium~II as the 
vortex core size $a_0$ is much smaller than any other 
characteristic length scale of the flow. However, during the time
evolution of a turbulent vortex tangle, there are important events
such as vortex reconnections, during which
this assumption breaks down. We shall discuss how these
events can be accounted for in the VFM at the end
of the section. 

\begin{enumerate}[label=(\roman*)]

\item{Lagrangian discretization}
%\subsubsection{Vortex spatial discretisation}

At each time, we discretize the vortex tangle $\cal L$ in $N_p$
Lagrangian vortex points
$\displaystyle\left \{ \mathbf{s}_i(t) \right \}_{i=1,\dots, N_p}\;$ 
The distance between neighbouring points is kept in the range 
$[ \delta/2, \delta ]$ by removing or inserting additional
points \cite{baggaley-barenghi-2011c}. As the vortex configuration 
evolves, so does this Lagrangian discretisation - $N_p$ depends
on the total length of the tangle. The vortex points evolve in time
according to Eq.~(\ref{eq:vortex_motion_final}) written in
dimensionless form, namely

\begin{equation}
\dot{\mathbf{s}_i} (t)= \mathbf{v}_{s_{\perp}}(\mathbf{s}_i,t)
+ \beta \; \mathbf{s}_i'(t) \times  \left ( \mathbf{v}_n(\mathbf{s}_i,t) - \mathbf{v}_s(\mathbf{s}_i,t) \right )
+\beta' \; \mathbf{s}_i'(t) \times \left \{ \mathbf{s}_i'(t) \times \left [ \mathbf{v}_n(\mathbf{s}_i,t) - \mathbf{v}_s(\mathbf{s}_i,t) \right ] \right \} \; \; ,
%\dot{\mathbf{s}_i} (t) = - \mathbf{s}_i'(t) \times \left \{ \mathbf{s}_i'(t) \times 
%\left [ \left ( 1 + \alpha \right ) \mathbf{v}_s(\mathbf{s}_i,t) - \alpha  \mathbf{v}_n(\mathbf{s}_i,t) \right ] \right \} 
%+ \beta \mathbf{s}'_i (t)\times  \left [ \mathbf{v}_n(\mathbf{s}_i,t) - \mathbf{v}_s(\mathbf{s}_i,t) \right ]       \; \; ,
\label{eq:vortex_motion_discrete}
\end{equation}
where spatial derivatives along the vortex lines are performed 
employing fourth-order finite difference schemes which account for the
varying mesh size along the vortex lines 
\cite{gamet-etal-1999,baggaley-barenghi-2011c},
and time integration is performed 
using the third-order Adams-Bashforth method.

\noindent
The interpolation of the normal fluid velocity $\mathbf{v}_n$ 
on each vortex point $\mathbf{s}_i$ presents some numerical issues;
the different schemes which we have tested
are outlined and discussed in Section \ref{subsec:interp}. 
The evaluation of the superfluid velocity $\mathbf{v}_s$ on the 
vortex points $\mathbf{s}_i$ via Eq. \refeq{eq:vs_no_dim} must
be dealt with caution as
the Biot-Savart integral diverges 
%for $\mathbf{x}=\mathbf{s}_i$ 
when $\mathbf{x} \rightarrow \mathbf{s}_i$ in Eq.~\refeq{eq:vs_no_dim}.
This singularity is removed in a standard fashion by splitting 
the Biot-Savart integral into local and nonlocal contributions 
\cite{schwarz-1985}, that is
(omitting time dependency to ease notation) by writing

\begin{equation}
\displaystyle
\mathbf{v}_s (\mathbf{s}_i)= \nabla \phi (\mathbf{s}_i) + 
\frac{\kappa}{4\pi}\ln \left ( \frac{\sqrt{\delta_i \delta_{i+1}}}{a_0} \right )  \mathbf{s}_i' \times \mathbf{s}_i''+ 
\frac{\kappa}{4\pi } \oint_{\mathcal{L}'}\frac{\mathbf{s}'(\xi)\times (\mathbf{s}_i-\mathbf{s}(\xi))}{|\mathbf{s}_i-\mathbf{s}(\xi)|^3} d\xi \;\;\;\; ,\label{eq:BS_desing}
\end{equation}

\noindent
where $\delta_i$ and $\delta_{i+1}$ are the lengths of the segments 
$[\mathbf{s}_{i-1}\, ,\,\mathbf{s}_i]$ and 
$[\mathbf{s}_{i}\, ,\,\mathbf{s}_{i+1}]$ respectively,
$\displaystyle \mathbf{s}_i'' 
= \partial^2 \mathbf{s}(\xi,t)/\partial \xi^2$ evaluated at
${\mathbf{s}_i}$ 
is the normal vector to the curve in $\mathbf{s}(\xi,t)=\mathbf{s}_i$,
and $\mathcal{L}'$ is the vortex configuration without 
the section between $\mathbf{s}_{i-1}$ and $\mathbf{s}_{i+1}$.
To match the periodic boundaries used in the integration of the 
Navier--Stokes equations for the normal fluid (Section \ref{subsec:NS}), 
we introduce periodic wrapping into the VFM. 
This involves creating copies of the vortex
configuration around the original configuration;
the contributions of these copies are 
included in the Biot-Savart integrals, Eq. \refeq{eq:BS_desing}. 
In order to speed-up the calculation of Biot Savart integrals for
dense tangles of vortices, Eq. \refeq{eq:BS_desing} is approximated
using a tree algorithm \cite{baggaley-barenghi-2012} which scales 
as $N_p \log N_p$ rather than $N_p^2$.

\item{Vortex reconnections}
%\label{subsubsec:reconnections}

In the limit of zero temperature, Equation \refeq{eq:vs} 
%describing the behaviour of the superfluid velocity governed by the quantised vortex lines 
expresses the superfluid's underlying incompressible, 
inviscid Euler dynamics 
in integral form \cite{saffman-1992}. Hence, on the basis
of Helmholtz theorem, the topology of the superflow 
ought to be frozen, \textit{i.e.} reconnections of vortex lines
are not envisaged.
We know however from experiments 
\cite{bewley-etal-2008,serafini-etal-2017} and from more microscopic models 
\cite{koplik-levine-1993,tebbs-etal-2011,kerr-2011,kursa-etal-2011,villois-etal-2017,galantucci-etal-2019} 
that when vortex lines come sufficiently close to each other, they
reconnect, exchanging strands,
as envisaged by Feynman \cite{feynman-1955}. 

In order to model vortex reconnections within the VFM,
we supplement Eq.~\refeq{eq:vortex_motion_discrete} 
with an \textit{ad-hoc} algorithmical reconnection procedure. This strategy was originally proposed by Schwarz \cite{schwarz-1978}, 
and since then a number of alternative algorithms have been proposed.
Whilst this procedure is essentially arbitrary, a number of
different implementations have been extensively tested and compared
\cite{baggaley-2012b}, finding no significant physical difference
between these implementations.

\end{enumerate}

%%% NS %%%%%%%%%%%%%%%

\subsection{Navier Stokes Solver}
\label{subsec:NS}

We solve Eq.\eqref{Eq:SelfConstModelEqVn} using a standard 
pseudo-spectral code in a three dimensional periodic domain of 
size $L_x\times L_y\times L_z$ with $n_x, n_y$ and $n_z$ 
collocation points in each direction. Derivatives of the fields 
are directly computed in spectral space, whereas non-linear terms 
are evaluated in physical space. The code is de-aliased using the 
standard 2/3-rule, and therefore the maximum wavenumber 
is $k_{\rm max}=2\pi \min{[n_x/L_x,n_y/L_y,n_z/L_z]}/3$. 
The main advantage of pseudo-spectral codes is that non-linear partial 
differential equations are solved with spectral accuracy; this means 
that spatial approximation errors decrease exponentially with the number 
of collocation points. The drawbacks of pseudo-spectral codes
are that the computational box must be periodic and 
Fourier transforms are intensively used. 

The external forcing in Eq.\eqref{Eq:SelfConstModelEqVn} consists 
of a superposition of random Fourier modes:

\begin{equation}
{\bf F}_{\rm ext}({\bf x})=\frac{f_0}{{\mathcal N}_f}\sum_{k_{\rm inf}\le |{\bf k}|\le k_{\rm sup}} {\bf F}_{\bf k}e^{i{\bf k}\cdot{\bf x}},
\end{equation}

\noindent
where ${\bf F}_{\bf k}$ is a Gaussian vector of zero mean 
and unit variance satisfying ${\bf F}_{\bf k}\cdot {\bf k}=0$ 
(incompressibility condition),
and ${\bf F}_{\bf k}^*={\bf F}_{-\bf k}$. ${\mathcal N}_f$ 
is a normalisation constant to impose that 
$\langle |{\bf F}_{\rm ext}({\bf x})|^2\rangle=f_0^2$. 
A second choice of forcing is the so-called \emph{frozen} forcing. 
It is simply obtained (after each time step)
by setting  the velocity field equal to a prescribe field for wave-vectors 
in a predefined range. This second forcing mimics a physical 
forcing working at constant velocity.

The pressure in the Navier--Stokes equation ensures the 
incompressible condition. The condition is easily satisfied
by inverting the Poisson equation 

\begin{equation}
-\frac{1}{\rho_n}\nabla^2 p = \nabla\cdot\left( \left ( \mathbf{v}_n \cdot \nabla \right ) \mathbf{v}_n\right)-
\frac{1}{\rho_n}\nabla\cdot\left( \oint_\mathcal{L} \delta(\mathbf{x-s}) \mathbf{f}_{ns}(\mathbf{s}) \mathrm{d}\xi\right).
\end{equation}

\noindent
Let us denote by $\mathcal{P}$ the projector into the 
subspace of divergence-free functions, and define
$P_{ij}=\delta_{ij}-\frac{\partial_i\partial_j}{\nabla^2}$
where $\partial_i=\partial/\partial x_i$. The equation of motion 
for the normal fluid simplifies to

\begin{equation}
\frac{\partial \mathbf{v}_n}{\partial t} + \mathcal{P}\left ( \mathbf{v}_n \cdot \nabla \right ) \mathbf{v}_n  =  \nu_n^0 \nabla^2 \mathbf{v}_n +\frac{1}{\rho_n}\mathcal{P}\oint_\mathcal{L} \delta(\mathbf{x-s}) \mathbf{f}_{ns}(\mathbf{s}) \mathrm{d}\xi+\frac{1}{\rho_n}\mathbf{F}_{\rm ext}. \label{Eq:NSprojected}
\end{equation}

\noindent
In practice, since the projector $\mathcal{P}$ takes a trivial form 
in Fourier space, we directly solve Eq.\eqref{Eq:NSprojected}.
Note that the force acting on the normal fluid due to the interaction 
with the vortex lines also needs to be projected.
Finally, we remark that the mean normal fluid velocity evolves
in time due to the friction force as a consequence of momentum 
transfer between components:

\begin{equation}
\frac{\partial \langle\mathbf{v}_n\rangle}{\partial t}  = \frac{1}{L_xL_yL_z}\frac{1}{\rho_n}\oint_\mathcal{L}  \mathbf{f}_{ns}(\mathbf{s}) \mathrm{d}\xi. \label{Eq:meanVn}
\end{equation}

%%% INTERPOLATION %%%%%%%%%%%%%%%

\subsection{Interpolation}
\label{subsec:interp}

The equation of motion of the vortex lines, 
Eq.~\eqref{Eq:SelfConstModelEqVFM}, needs the values of the normal 
fluid velocity at the Lagrangian discretization points along the
vortex lines, where ${\bf v}_n$ is not known. In principle, 
as the normal fluid is periodic, any inter mesh value can be computed 
exactly (within spectral accuracy) by using a Fourier transform. 
We call this interpolation scheme \emph{Fourier interpolation} and use it
as benchmark for comparisons. The Fourier interpolation is extremely 
costly and prohibitive for practical applications, as it requires 
$n_x n_yn_z$ evaluations of complex exponentials for each 
Lagrangian point along the vortex lines.
Affordable interpolation schemes are typically defined on physical 
space and their accuracy depends on the order of the method
and on the regularity of the field. In three dimensions,
the most commonly used schemes 
are \emph{Nearest neighbours}, where the value of the closest grid 
point is taken, and \emph{Tri-Linear} or
\emph{Tri-Cubic} interpolation,
performed in each direction by a line or a cubic polynomial. 
More recently, a new approach based on fourth-order \emph{B-splines} 
has been proved to be specially well adapted to pseudo-spectral codes,
and found to be non-expensive and highly accurate \cite{vanHinsberg2012}.

In order to study the effect of different interpolation schemes, we study 
how a superfluid vortex ring moves in a static spatially dependent flow. 
We do not take into account yet the retroaction of the vortex filament 
on the normal fluid which we keep fixed. We consider a domain of 
size $L_x=L_y=L_z=2\pi$; for the normal fluid we choose
the following superposition of ABC flows at different scales:

\begin{equation}
{\bf v}_n(x,y,z)=\sum_{n=1}^{n_{\rm max}}(B\cos{n y}+C\sin{nz},A\sin{x}+C\cos{nz},A \cos{nx}+B\sin{ny}),\label{Eq:ABCFlow}
\end{equation}

\noindent
with $A=1,B=-1,C=5$ and $n_{\rm max}=10$. We evaluate the field on
a mesh with $N=128$ and $256$ collocation points in each direction. 
As the largest wavevector is $\sqrt{3}\,n_{\rm max}$, the Fourier 
interpolation is exact (up to round-up errors).

\begin{figure}[h]
\includegraphics[width=0.48\textwidth]{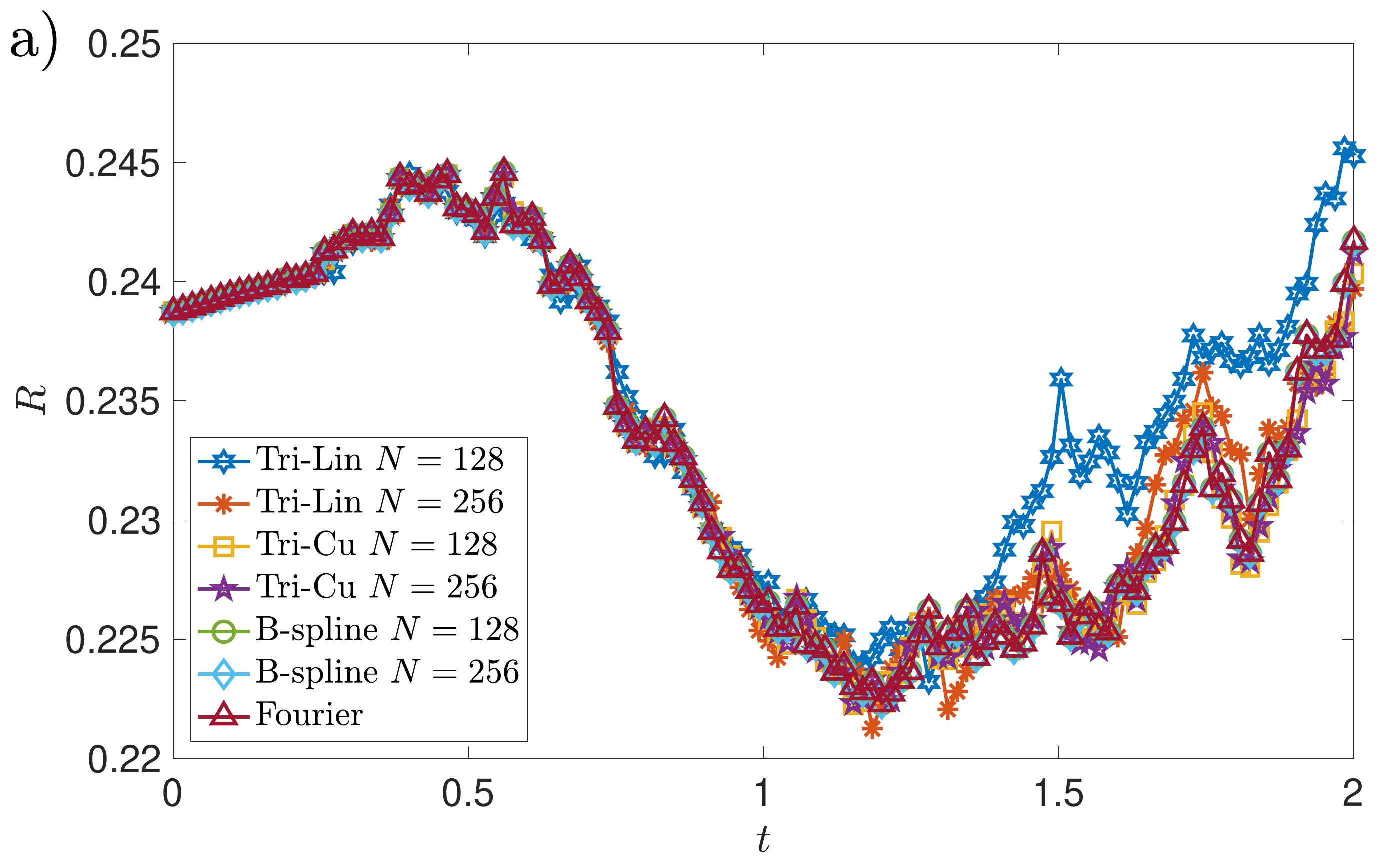} 
\includegraphics[width=0.48\textwidth]{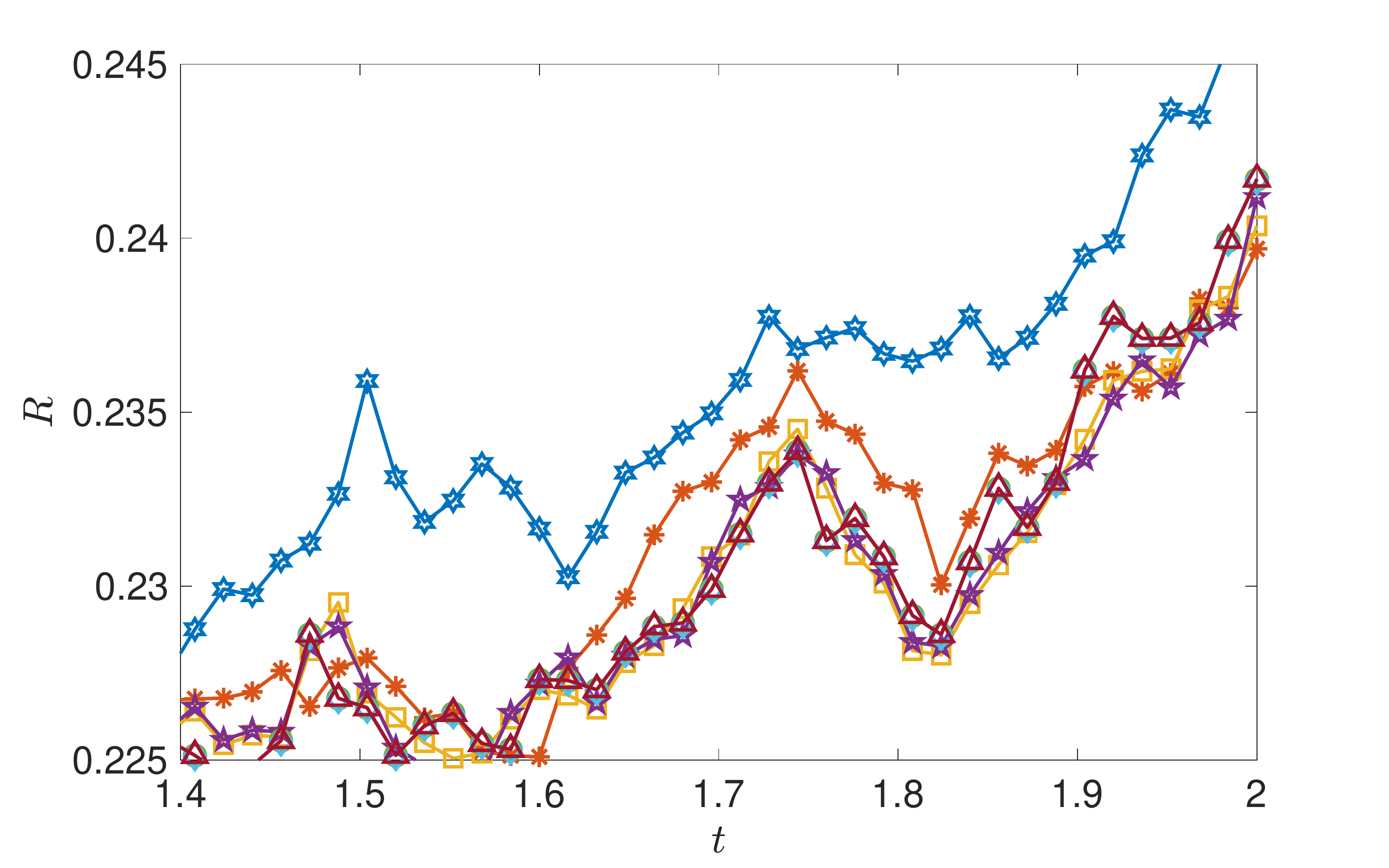} 
\caption{(Color online) Temporal evolution of the average radius of a vortex ring 
for different interpolation schemes and normal fluid grid resolutions 
(arbitrary units). The right figure displays an enlargement
of the left figure at late times to better judge the interpolation schemes.}
\label{Fig:RingRadiusEvolInterp}
\end{figure}

We initialise a vortex ring of size $R=0.2387$ (in non-dimensional units)
and set the temperature at $T=1.95 \rm ~K$. We let the ring evolve with the static normal fluid 
in the background. As the ring evolves, it deforms in the presence
of the highly non-homogeneous normal fluid. In order to obtain a 
quantitative comparison between different schemes, we measure the average 
radius of the voretx ring as a function of time 
$R(t)=\langle |{\bf s(\xi)}-{\bf s}_{\rm center}|\rangle$ 
where ${\bf s}_{\rm center}=\langle {\bf s}\rangle$ and the average 
is performed over vortex points. The temporal evolution of $R(t)$ 
is displayed in Fig.\ref{Fig:RingRadiusEvolInterp} for two different 
resolutions and different interpolation schemes.

\noindent
As expected, it is apparent that the error decreases when the number 
of collocation points of the normal fluid grid increases. It is also clear
that Tri-linear interpolation gives  poor results.

\begin{figure}[h]
\begin{center}
\includegraphics[width=0.8\textwidth]{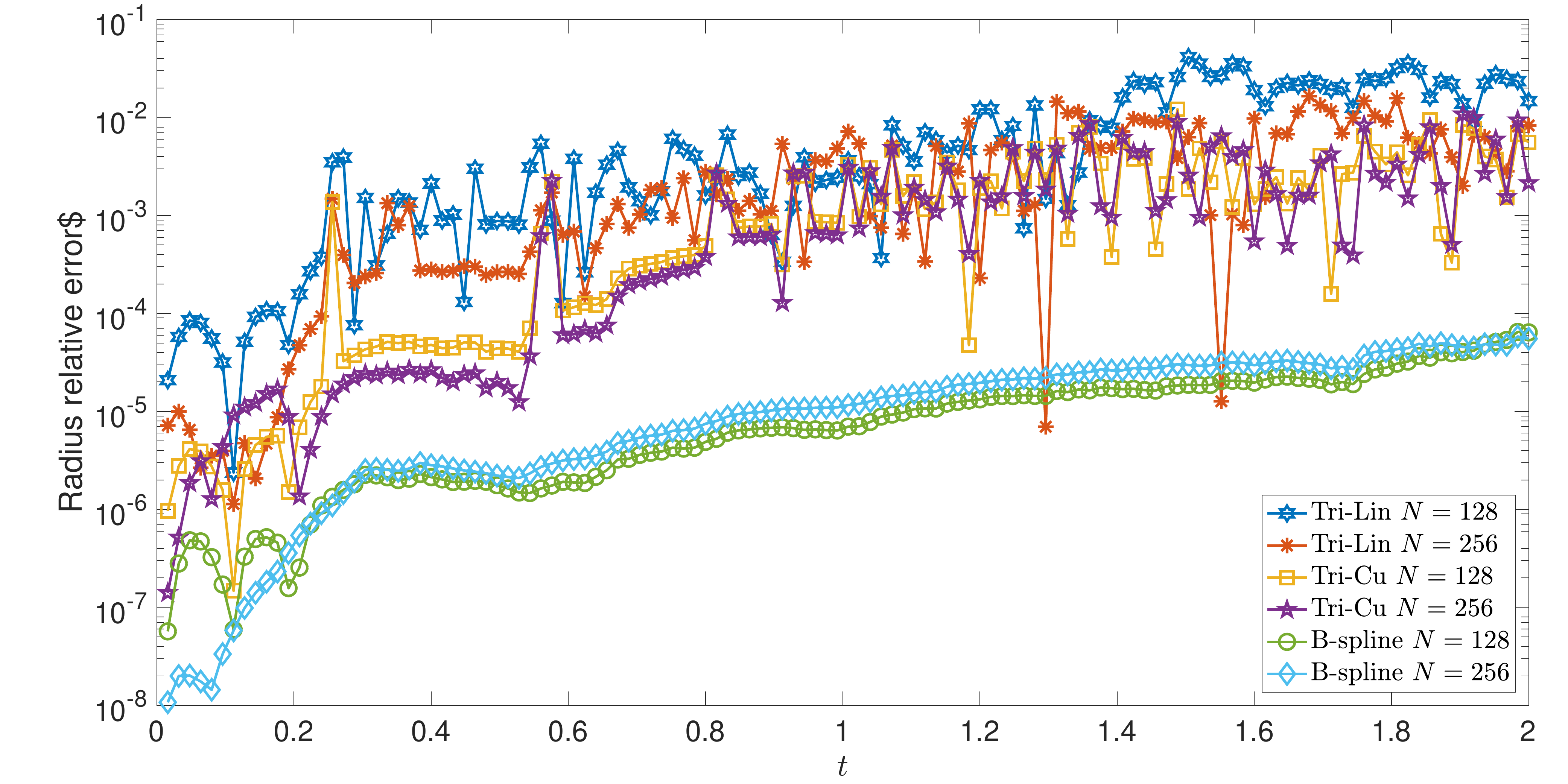} 
\caption{(Color online). Temporal evolution of the relative error $|R(t)-R_{\rm ref}(t)|/R_{\rm ref}(t)$, where the reference evolution is obtained with Fourier interpolation (arbitrary units).}
\label{Fig:RingRadiusEvolInterpError}
\end{center}
\end{figure}

To better quantify the accuracy of the interpolation methods, we compute 
the relative error of the radius with respect to the reference
radius obtained by Fourier interpolation 
$|R(t)-R_{\rm ref}(t)|/R_{\rm ref}(t)$. 
The time evolution of this relative error
is displayed in Fig.\ref{Fig:RingRadiusEvolInterpError}.
Remarkably, the B-spline interpolation reduces
the interpolating errors  considerably. 
The extra cost of the B-spline scheme is just one single fast Fourier 
transform independently the number of points to be interpolated, and
this is why we choose it for our local self-consistent approach, \textit{FOUCAULT}.

%%% EXTRAPOLATION %%%%%%%%%%%%%%%

\subsection{Friction distribution}
\label{subsec:extrap}

\begin{figure}[h]
\begin{center}
\includegraphics[width=0.4\textwidth]{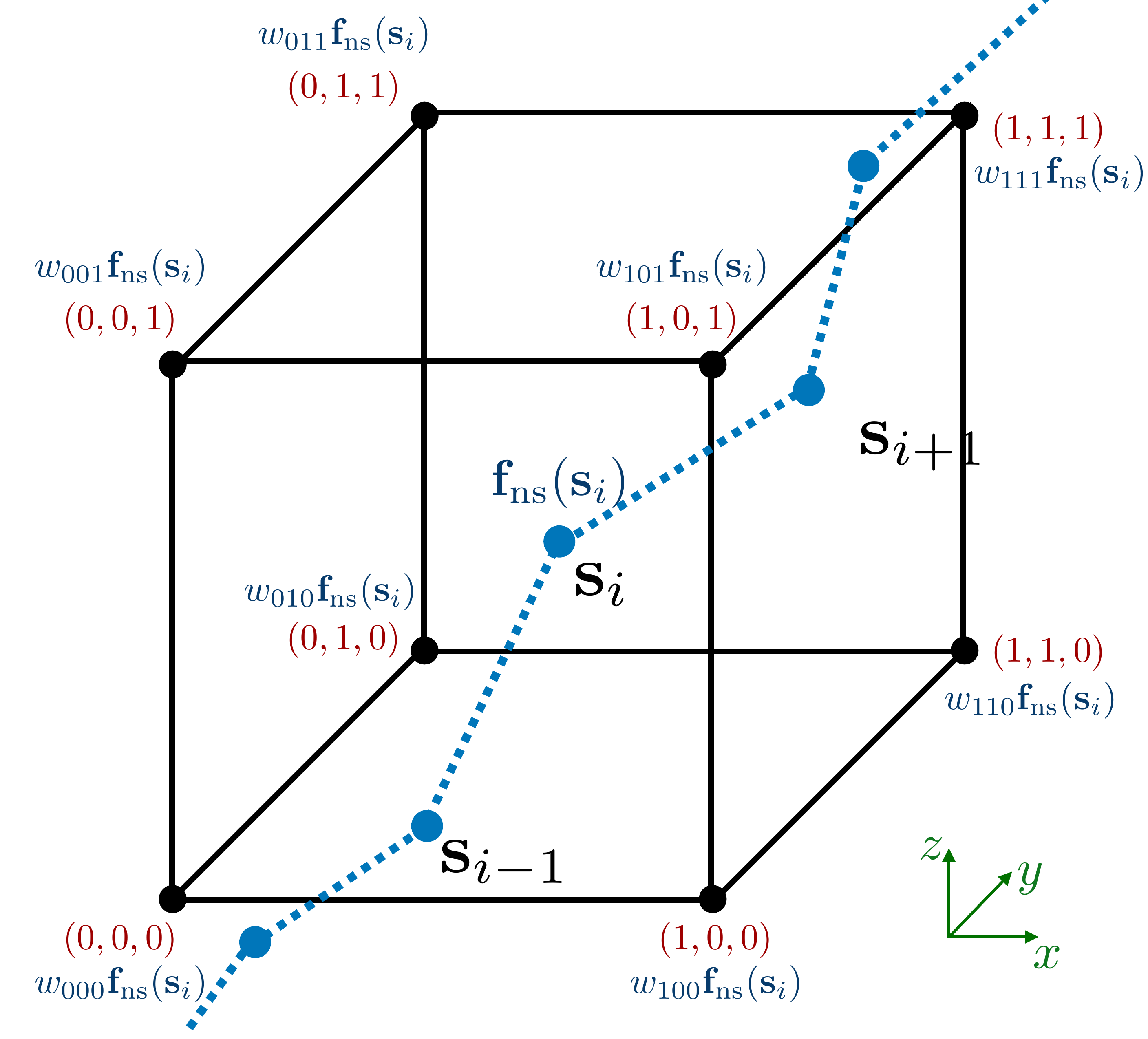} 
\caption{Sketch of the force distribution and weights.}
\label{Fig:sketchweights}
\end{center}
\end{figure}

We now discuss the back-reaction force of the vortex lines on 
the  normal fluid. 
This force, seen from the normal fluid, is $\delta$-supported on the 
$N_p$ Lagrangian points along the vortex lines
$\displaystyle\left \{ \mathbf{s}_i(t) \right \}_{i=1,\dots, N_p}\;$
%Theoretically, as the Navier--Stokes solver is pseudo-spectral, the retro-action force could be computed exactly in Fourier space as follows
%\begin{equation}
%\mathcal{P}\oint_\mathcal{L} e^{i{\bf k}\cdot \mathbf{s}} \mathbf{f}_{ns}(\mathbf{s}) \mathrm{d}|\mathbf{s}|.
%\end{equation}
%However, such approach is way too expensive...
and needs to be distributed on the grid points where the normal fluid is 
computed. More precisely, we denote by $(\zeta,\mu,\chi)$, 
with ${\zeta,\mu,\chi}\in [0,1)$, the neighboring grid points 
of ${\bf s}_i(t)$. The force ${\bf f}_{\rm ns}({\bf s}_i(t))$ 
exerted by the vortex line on the normal fluid has to be 
distributed among neighbouring points using 
weights $w_{\zeta,\mu,\chi}$, as showed schematically in
Fig.\ref{Fig:sketchweights}.

By definition, the weights satisfy 
$\displaystyle\sum_{\zeta,\mu,\chi=0}^1 w_{\zeta,\mu,\chi}=1$. 
In principle, extra smoothing of the force field can be applied after
the force has been distributed.
This numerical problem is often faced in active matter systems, 
where small point-like particles ({\it i.e.} swimmers, plankton, bacteria) 
retro-act on a classical turbulent flow. Let us rewrite the normal fluid 
equation considering the discretisation of the vortex lines discussed 
in Section \ref{subsec:VFM}:

\begin{equation}
\displaystyle
\frac{\partial \mathbf{v}_n}{\partial t} + \mathcal{P}\left ( \mathbf{v}_n \cdot \nabla \right ) \mathbf{v}_n  =  \nu_n^0 \nabla^2 \mathbf{v}_n +
\frac{1}{\rho_n}\sum_{i=1}^{N_p}\mathcal{P}\delta(\mathbf{x}-\mathbf{s}_i(t)) \mathbf{f}_{ns}^{\,i}(t)\delta_i \;\; , \label{Eq:NSVortexDiscrete}
\end{equation}

\noindent
where $\mathbf{f}_{ns}^i(t):= {\bf f}_{\rm ns}({\bf s}_i(t))$, $\delta_i$ 
is the length of the $i$-th vortex line element and the
dependence on time of the friction force is explicitly showed,
as it will play an important role in the subsequent part of the discussion. 
For the sake of simplicity, the external force has been omitted 
in Eq.~(\ref{Eq:NSVortexDiscrete}) 
and a simple Riemann sum has been used to approximate the line integral. 
Note that a trapezoidal rule, which is a better approximation, 
may be numerically used by the simple replacement
$\delta_i \to (\delta_i+\delta_{i+1})/2$ and readjusting 
the indices of the sum. 
It is thus evident that our problem is formally identical to 
coupling discrete, point-like, active particles to the turbulent 
flow of a classical viscous fluid. 

\begin{figure}[h]
\begin{center}
\includegraphics[width=0.8\textwidth]{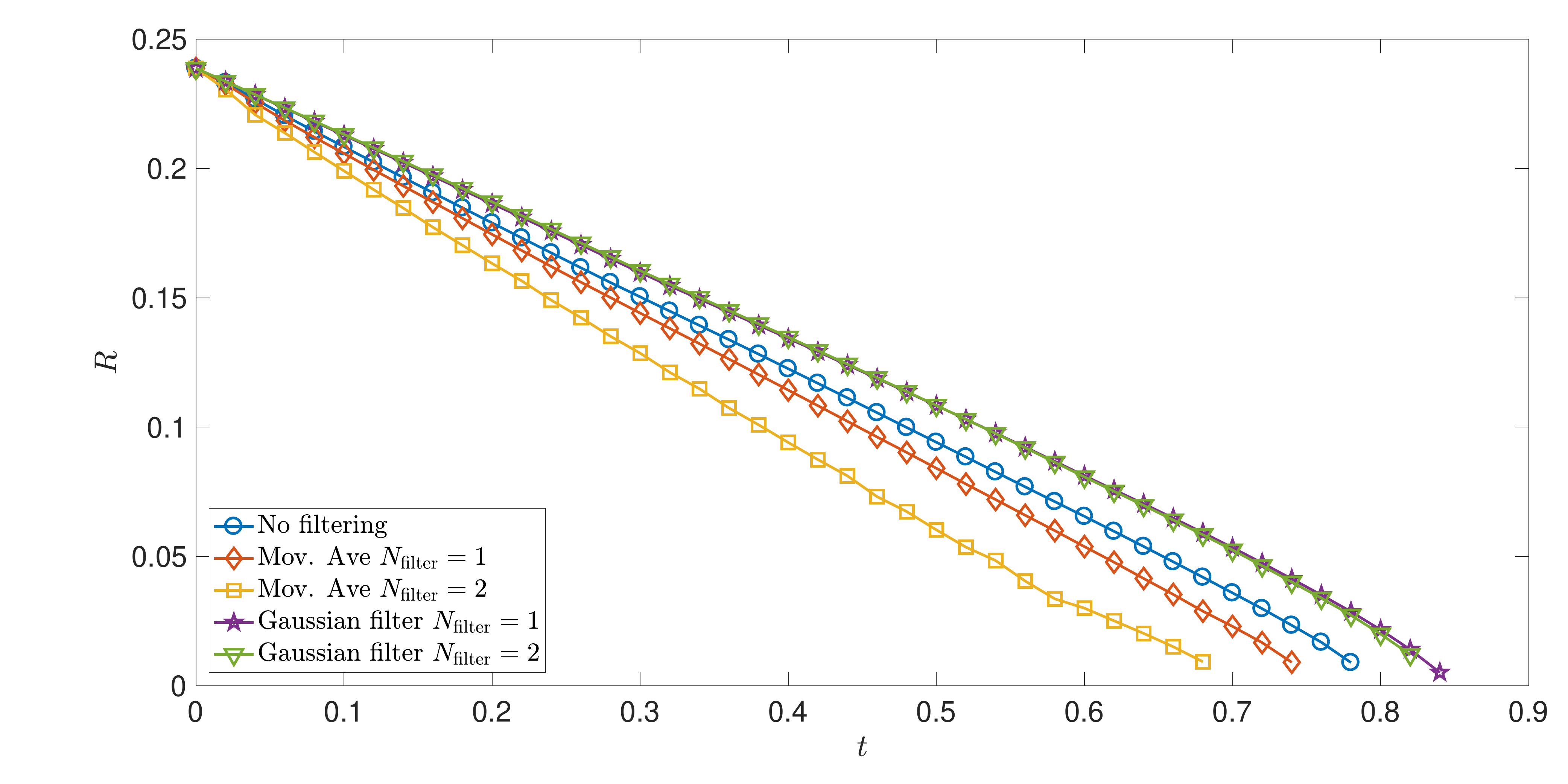} 
\caption{(Color online). Temporal evolution of the radius of a vortex 
ring with in an initially uniform, quiescent normal fluid.  
The force acting on the normal fluid is distributed to the nearest 
neighbour point of the mesh. Different filtering procedures are tested.}
\label{Fig:RingRadiusEvolReDitribForce}
\end{center}
\end{figure}

It is well known that in problems of active matter 
properties such as \textit{e.g.} aggregation and condensation, mixing or/and
growing of species may strongly depend on the choice of the 
force distribution method and the filtering scale  
\cite{gualtieri_picano_sardina_casciola_2015,GualtieriPREActiveParticlesPRE}. 
The same issues appear in our self-consistent local model \textit{FOUCAULT}. 
In order to illustrate this problem, we consider the simple case of 
a vortex ring moving in a initially quiescent normal fluid, 
similar to the setting studied in 
\cite{kivotides-barenghi-samuels-2000}. We consider a ring of radius 
$R=0.2387$ and set the temperature at $T=1.95^\circ$K. 
Because of friction, we expect the ring to shrink. 
We compare the temporal evolution of its radius 
employing different filtering methods. We distribute the force to the 
nearest neighbour grid points, i.e. $w_{\zeta,\mu,\chi}=0$ 
for all grid points 
except the nearest one to ${\bf s}_i$. The resulting force field 
is then filtered by using either a moving average over $N_{\rm filter}$ 
points or a Gaussian kernel of width $N_{\rm filter}\Delta x$, 
where $\Delta x=L_x/N_x$ is the mesh size.
Figure \ref{Fig:RingRadiusEvolReDitribForce} displays the 
temporal evolution of the vortex ring radius for the different 
schemes. It is clear that the shrinking rate of the radius of the ring
depends strongly on the filtering procedure
which is employed.

\noindent
This dependence is clearly spurious. A numerical method
based on physics principles is hence needed. Since our context is turbulence,
it makes sense to adopt the same
rigorous regularisation approach which has been used to take into account
the strongly localised response of point-like 
particles in classical turbulence
\cite{gualtieri_picano_sardina_casciola_2015,GualtieriPREActiveParticlesPRE}. 
The advantage of adopting this method is that the regularisation of the 
exchange of momentum between point-like particles and viscous flows 
is based on the physics of the generation of vorticity 
and its viscous diffusion at very small scales. 
In our case, the justification of this method arises from
the very small Reynolds numbers of the flow based on 
the vortex core radius and the small velocity of the vortex line with respect
to the normal fluid.

We refer the reader to the original papers
\cite{gualtieri_picano_sardina_casciola_2015,GualtieriPREActiveParticlesPRE} 
for further details; in brief, this approach which we borrow from 
classical turbulence is based on the solution of the delta-forced,
linear unsteady Stokes equation. Accordingly
to the classical formulation, we introduce the time-delay $\epsilon_R$ 
coinciding with the time interval during which the localised vorticity
(generated by the relative motion between the vortex line and the
normal fluid) is diffused to the relevant 
hydrodynamic scale of the flow, the spacing $\Delta x$ of 
the computational grid which the normal fluid velocity field is calculated on. 
This finite time delay, $\epsilon_R$, regularises the delta-shaped 
nature of the friction force 
$\mathbf{f}_{ns}^{\,i}(t)\delta(\mathbf{x}-\mathbf{s}_i(t))$ 
in a natural way
via the fundamental solution of the diffusion equation:

\begin{equation}
\displaystyle
g[\mathbf{x}-\mathbf{s}_i(t-\epsilon_R),\epsilon_R]=\frac{1}{(4\pi\nu\epsilon_R)^{3/2}}
\exp\left [ -\frac{|| \mathbf{x}-\mathbf{s}_i(t-\epsilon_R) ||^2}
{4\nu\epsilon_R}\right ] \;\; , 
\label{eq:Gaussian.kernel}
\end{equation}

\noindent
This solution, Eq.~\refeq{eq:Gaussian.kernel}, is a Gaussian with 
standard deviation $\sigma_R=\sqrt{2\nu\epsilon_R}$. 
The resulting expression for the friction force exerted by the $i$-th 
vortex line element on the normal fluid at the time $t$
at the point $\mathbf{x}$ is 
$\mathbf{f}_{ns}^{\,i}(t-\epsilon_R)g[\mathbf{x}-\mathbf{s}_i(t-\epsilon_R),\epsilon_R] \delta_i$
\cite{gualtieri_picano_sardina_casciola_2015,GualtieriPREActiveParticlesPRE},
yielding the following modified Navier--Stokes equation
for the normal fluid velocity field:

\begin{equation}
\displaystyle
\frac{\partial \mathbf{v}_n}{\partial t} + \mathcal{P}\left ( \mathbf{v}_n \cdot \nabla \right ) \mathbf{v}_n  =  \nu_n^0 \nabla^2 \mathbf{v}_n +
\frac{1}{\rho_n}\sum_{i=1}^{N_p}\mathcal{P}\;\mathbf{f}_{ns}^{\,i}(t-\epsilon_R)g[\mathbf{x}-\mathbf{s}_i(t-\epsilon_R),\epsilon_R]\delta_i \;\; . 
\label{Eq:NSdelayed}
\end{equation}

From the physical point of view, Eq.~(\ref{Eq:NSdelayed}) implies that 
the strongly localised 
vorticity injected in the normal flow by the relative motion of the vortices 
is neglected until it has been diffused by viscosity to a characteristic 
length scale $\sigma_R=\sqrt{2\nu\epsilon_R}$. 
To be consistent, and in order to take into account the vortex induced 
disturbances as soon as the relevant hydrodynamic scales are
affected, we choose the finite time delay $\epsilon_R$ so that 
$\sigma_R/\Delta x=1$. Extensive tests performed in the original paper
\cite{gualtieri_picano_sardina_casciola_2015} ensure that 
$\sigma_R/\Delta x=1$ is a suitable choice.

\begin{figure}
\begin{center}
\includegraphics[width=0.4\textwidth]{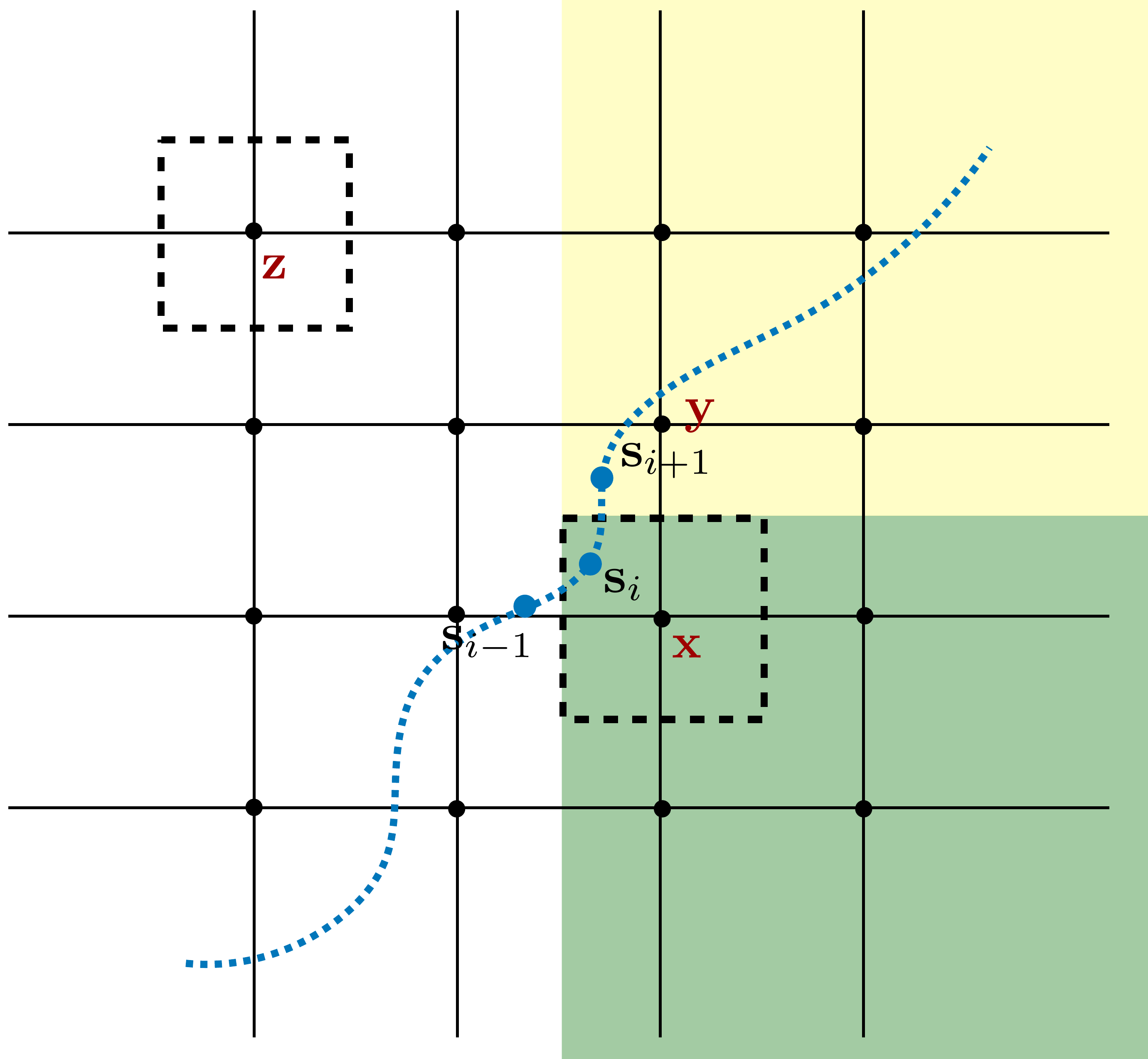} 
\caption{(Color online). Schematic two-dimensional 
distribution of the friction force.}
\label{Fig:sketch}
\end{center}
\end{figure}

In theory, for each $i$-th vortex line element, it is possible to 
compute the corresponding weight for each point of the numerical grid:
it would sufficient to integrate the Gaussian kernel 
$g[\mathbf{x}-\mathbf{s}_i(t-\epsilon_R),\epsilon_R]$ over the volume 
$\Delta V=\Delta x\Delta y\Delta z$ centred on the grid point. 
For instance, as displayed in the schematic
two-dimensional Fig.~\ref{Fig:sketch}, 
the force generated by the $i$-th vortex element at ${\bf s}_i$ 
contributes to the force computed on mesh point ${\bf x}$, 
but also on mesh point 
${\bf z}$, the weights being the integral of 
$g(\mathbf{x}-\mathbf{s}_i,\epsilon_R)$ over their respective cells, 
represented by black dashed squares in the Fig.~\ref{Fig:sketch}. 
This procedure would have to be repeated on each grid point for 
each vortex element.
Although exact, this approach would be extremely costly. 
In addition, referring again to Fig.\ref{Fig:sketch}, 
the weight corresponding to mesh point 
${\bf z}$ is very small as we choose $\sigma_R/\Delta x=1$; the diffusion 
based regularisation which we employ localises in fact the 
force in a sphere of radius $\sigma_R$ centred at ${\bf s}_i$. 
As a consequence, instead of distributing the force on each grid point
by computing $N_x\, N_y\, N_z\, N_p$ integrals 
$g(\mathbf{x}-\mathbf{s}_i,\epsilon_R)$ 
all over the grid, we distribute the force 
only to neighbouring mesh points of ${\bf s}_i$, taking care of 
including the weights of far grid points. Proceeding in this fashion,
the total force is preserved, as the integral of 
$\displaystyle \int_{\mathbb{R}^3} g[\mathbf{x}-\mathbf{s}_i(t-\epsilon_R),\epsilon_R]\mathrm{d}{\bf x}=1$.
As matter of example, in the sketch outlined in Fig.\ref{Fig:sketch}, 
the ${\bf f}_{\rm ns}({\bf s}_i)$ of point ${\bf s}_i$ 
will be distributed among only the four neighbouring points. 
In order to use a conservative scheme, for each of the neighbouring points
the Gaussian kernel $g(\mathbf{x}-\mathbf{s}_i,\epsilon_R)$ 
will be integrated on the corresponding full quadrant (octant in 
three dimensions): in 
the two-dimensional simplified sketch in Fig.\ref{Fig:sketch}, 
this corresponds to integrating $g$ over the region coloured in 
green for point  ${\bf x}$ 
and over the yellow region for point ${\bf y}$. 
The generalisation to three dimension is straightforward. 
The advantage of this approach is that the space integrals of $g$ 
may actually be computed analytically.

The computation of the integrals leads to the weights $w_{\zeta,\mu,\chi}$ 
that will be used to distribute the force among the neighbouring 
grid points of ${\bf s}_i$, as illustrated in Fig.\ref{Fig:sketchweights}. 
Note that because the Gaussian kernel $g$ is normalised to one, 
the requirement that
$\displaystyle \sum_{\zeta,\mu,\chi=0}^1 w_{\zeta,\mu,\chi}=1$ 
is satisfied by construction.
To compute the weights $w_{\zeta,\mu,\chi}$, we first note 
that the integrals of $g(\mathbf{x}-\mathbf{s}_i,\epsilon_R)$ 
can be factorised in each
Cartesian direction: we will therefore only compute the contribution 
in the $x$ direction, the computation of the contributions of the 
other directions
is formally identical. The weight for the grid point of the left 
of $s_i^x$ (indicated with $\lfloor s_i^x\rfloor$, 
corresponding to $\zeta=0$) 
results from the one-dimensional integral over 
$(-\infty,\lfloor s_i^x\rfloor+\frac{\Delta x}{2})$, 
whereas the weight for the grid point on the right
($\lfloor s_i^x\rfloor +\Delta x$, $\zeta=1$), 
stems from the integral over 
$(\lfloor s_i^x\rfloor+\frac{\Delta x}{2},\infty)$. 
The calculation of the one-dimensional weight is straightforward;
we have:

\begin{eqnarray}
&& w_\zeta[s_i^x]=\zeta+(1-2\zeta)\frac{1}{2}\left(1+Erf{\left[-\frac{\tilde{s}_i^x-\frac{1}{2}}{\sqrt{2}(\sigma_R/\Delta x)}\right]}\right)\\
&& \tilde{s}_i^x=\frac{s_i^x-\lfloor s_i^x\rfloor}{\Delta x}\in[0,1].
\end{eqnarray}
The integrals $w_\mu[s_i^y]$ and $w_\chi[s_i^z]$ over the two remaining directions are performed in the same fashion. 
The diffusion-based weights for the three dimensional grid are finally given by
\begin{equation}
w_{\zeta,\mu,\chi}=w_\zeta[s_i^x]w_\mu[s_i^y]w_\chi[s_i^z]\label{eq:finalweights}
\end{equation}

\subsection{Numerical Strategy and Parallelisation}
\label{subsec:parallel}

We have implemented the numerical integration of the fully-coupled 
local model taking advantage of modern parallel computing. 
The solvers of the VFM and the Navier--Stokes equations are 
of very different nature, but they need to interact only
through the evaluation of the friction force. In this first version 
of the solver, we have opted for an hybrid OpenMP-MPI parallelisation scheme. 
The two solvers are handled by different MPI processes. 
Each MPI process contains many OpenMP threads, so each solver is also 
independently parallelised by using this shared memory library. 
The evaluation of the friction force requires communication between the 
two solvers that do not have access to each other fields and variables. 
This communication is managed through the 
Message Passing Interface (MPI) library at the end of each time step. 
%and the friction is \red{then} \red{[NOTE: please check]} applied. 
This scheme is naturally adapted to modern clusters that contain many nodes, 
with each node having a large number of CPUs with shared memory. 

Finally, we remark that the values of the time step 
for the Navier--Stokes equations 
and for the VFM solver need not be the same; typically, the VFM
requires a smaller time step. In order to speed up the code, 
depending on the physical problem, each solver can perform 
sub-loops (in time) to ensure numerical stability and 
an efficient integration of the full model.

%%% EXAMPLES %%%%%%%%%%%%%%%

\section{Applications}
\label{sec:examples}

In this section we show two physical applications of our fully-coupled
local model \textit{FOUCAULT}: (i) a superfluid vortex ring moving (i) in an initially
stationary normal fluid and (ii) in a turbulent normal fluid.
Our aim is not to introduce any new physical phenomena 
but simply validate our approach and show its computational capabilities.

\begin{figure}[h!]
\begin{center}
\includegraphics[width=0.48\textwidth]{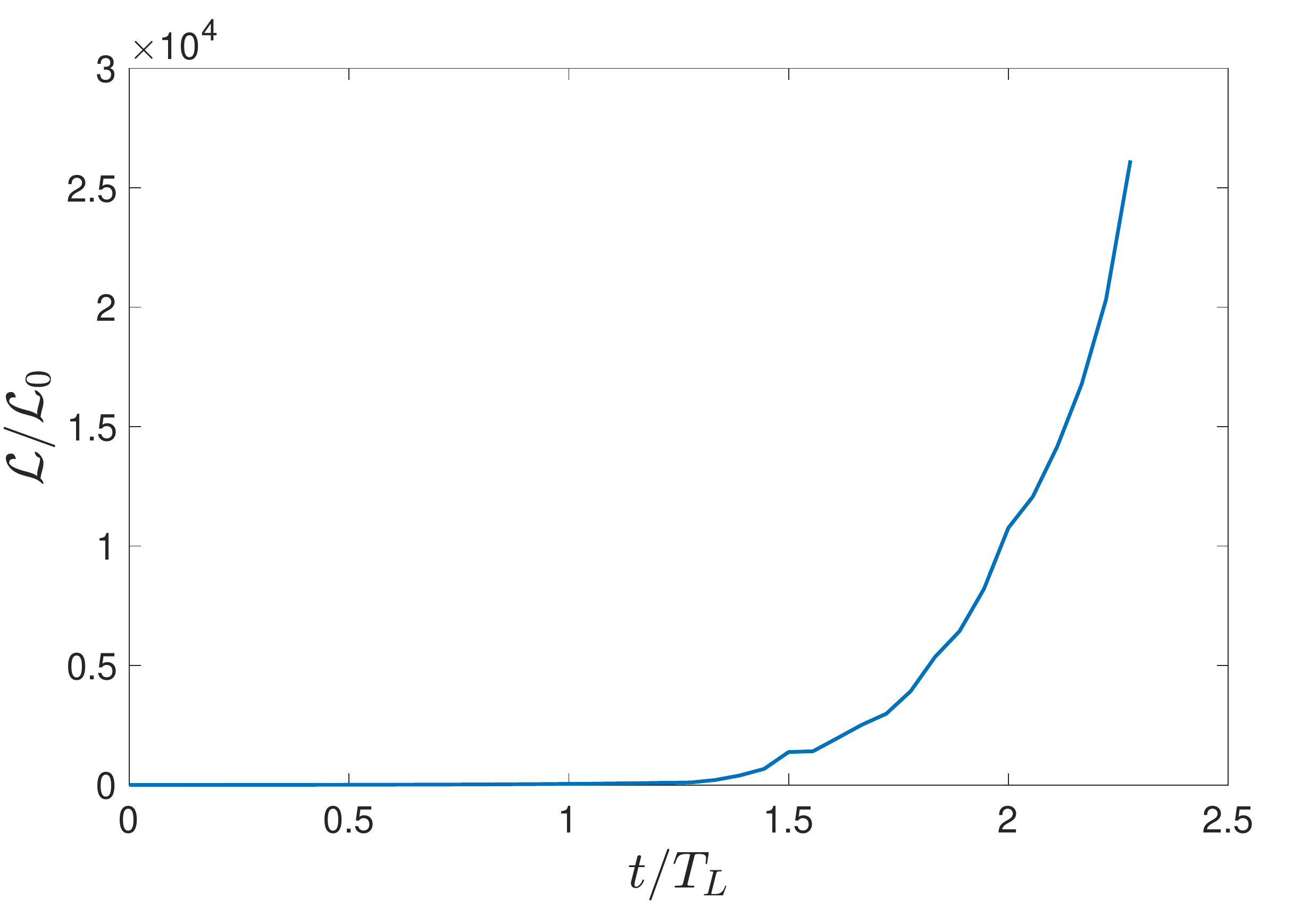} 
\includegraphics[width=0.48\textwidth]{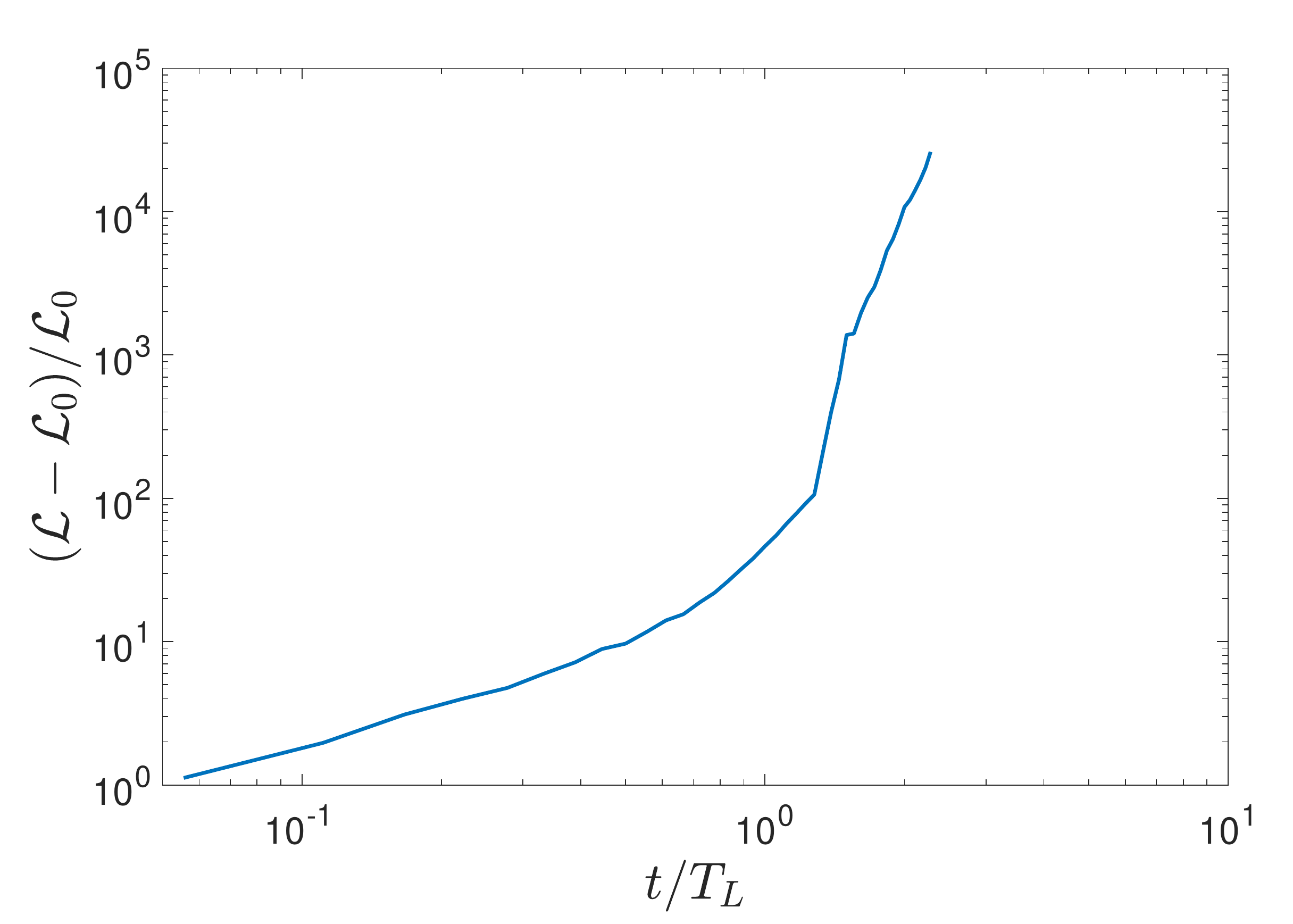} 
\caption{(Color online). Left panel: 
Temporal evolution of the total vortex length 
$\mathcal{L}$ resulting form a (initially perfect) vortex ring of 
length $\mathcal{L}_0=2\pi R \approx 1.5$ 
in a turbulent normal fluid background. 
Right panel: relative increase of the vortex length during the
initial stage.}
\label{Fig:VortexLengthRingInTurb}
\end{center}
\end{figure}

\begin{enumerate}[label=(\roman*)]

\item{Vortex ring in initially quiescent normal fluid}

We first study the vortex configuration investigated in the 
pioneer work of Kivotides et al. \cite{kivotides-barenghi-samuels-2000}:
a vortex ring moving in a initially quiescent normal fluid. They
observed that, due to the interaction between superfluid and normal
fluid, two concentric vortex rings are created
in the normal fluid, accompanying the superfluid vortex ring, forming
a triple vortex structure. We integrate our fully-coupled model
using as initial condition a superfluid vortex ring of 
radius $R=0.2387$ in a box of size $2\pi$ and set the temperature 
to $T=1.95 \rm~K$. 

As the superfluid vortex ring travels,
it sets in motion the surrounding normal fluid, thus losing energy
and shrinking. A visualisation of the resulting flow
is displayed in Fig.\ref{Fig:TripleRing}.
The superfluid vortex ring is displayed in green and the two normal fluid
vortex rings which are generated are rendered in orange-reddish colours,
forming the same triple vortex structure discovered by Kivotides et al.
\cite{kivotides-barenghi-samuels-2000}. 
On the plane perpendicular to the vortex rings we also show
the normal fluid translational velocity in the direction of propagation
of the three vortex rings, rendered in blue-white colours:
a wake (or jet) behind the superfluid ring, recently identified
\cite{mastracci-etal-2019}, is apparent.
%\red{[NOTE: the referee may ask us to do a more quantitative comparison with
%Mastracci]}

\begin{figure}[h]
\begin{center}
\includegraphics[width=0.785\textwidth]{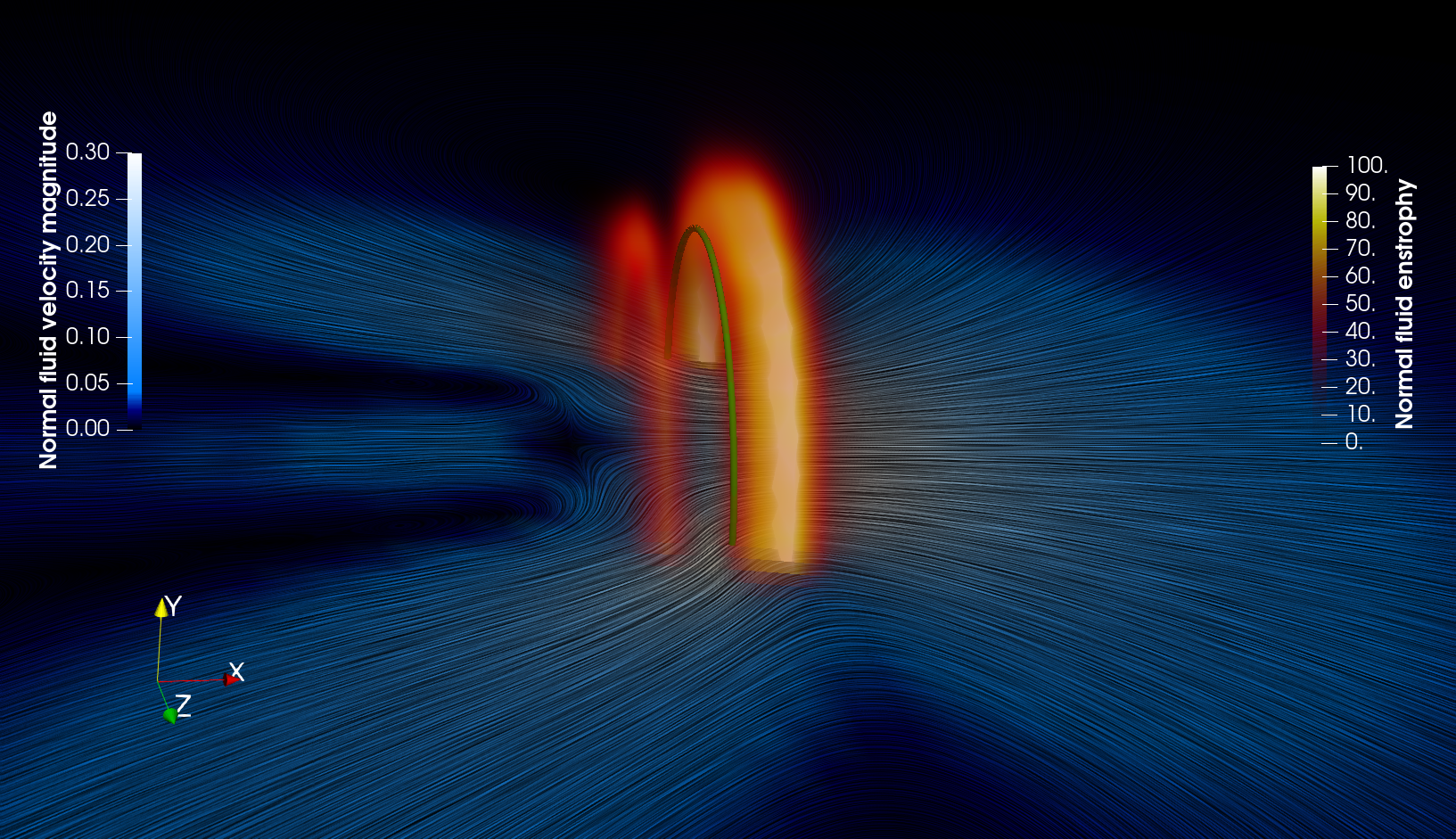} 
\caption{(Color online). A superfluid vortex ring moving
in the normal fluid initially at rest. Half of the superfluid vortex
ring is visible as a green line intersecting the xy plane; the superfluid
vortex ring moves to the right along the x direction.
The normal fluid enstrophy is displayed by the orange-reddish-black
rendering: two concentric normal fluid vortex rings are visible, slightly
ahead and slightly behind the superfluid vortex ring, 
travelling in the same direction. The normal fluid velocity magnitude is also displayed using
a black-blue-white rendering on the xy plane.
}
\label{Fig:TripleRing}
\end{center}
\end{figure}

\item{Vortex ring in turbulent normal fluid.}

Our second application is the effect of turbulence in the normal fluid
component on the dynamics of a superfluid vortex ring.
As discussed in section \ref{subsec:NS}, we can easily add an external 
forcing ${\bf F}_{\rm ext}$ to the Navier--Stokes equations in order to 
sustain the normal fluid turbulence. We first produce a turbulent 
state using a volumetric external forcing at large scales ($k_{\rm sup}=1$). We use 
a resolution of $256^3$ points, which allows us to obtain a Reynolds 
number based on the Taylor micro-scale equal to
$Re_\lambda=127$.
Once the turbulent state is prepared, we study the evolution at temperature $T=1.95 \rm ~K$ of a 
superfluid vortex ring of radius $R=35\eta$, where $\eta$ is the Kolmogorov length scale of the flow. The initial condition used in our 
self-consistent local model is displayed in 
Fig.~\ref{Fig:VizRingInTurb} (top left panel).

\begin{figure}[h]
%\begin{centering}
    \subfigure{\includegraphics[width=0.49\textwidth]{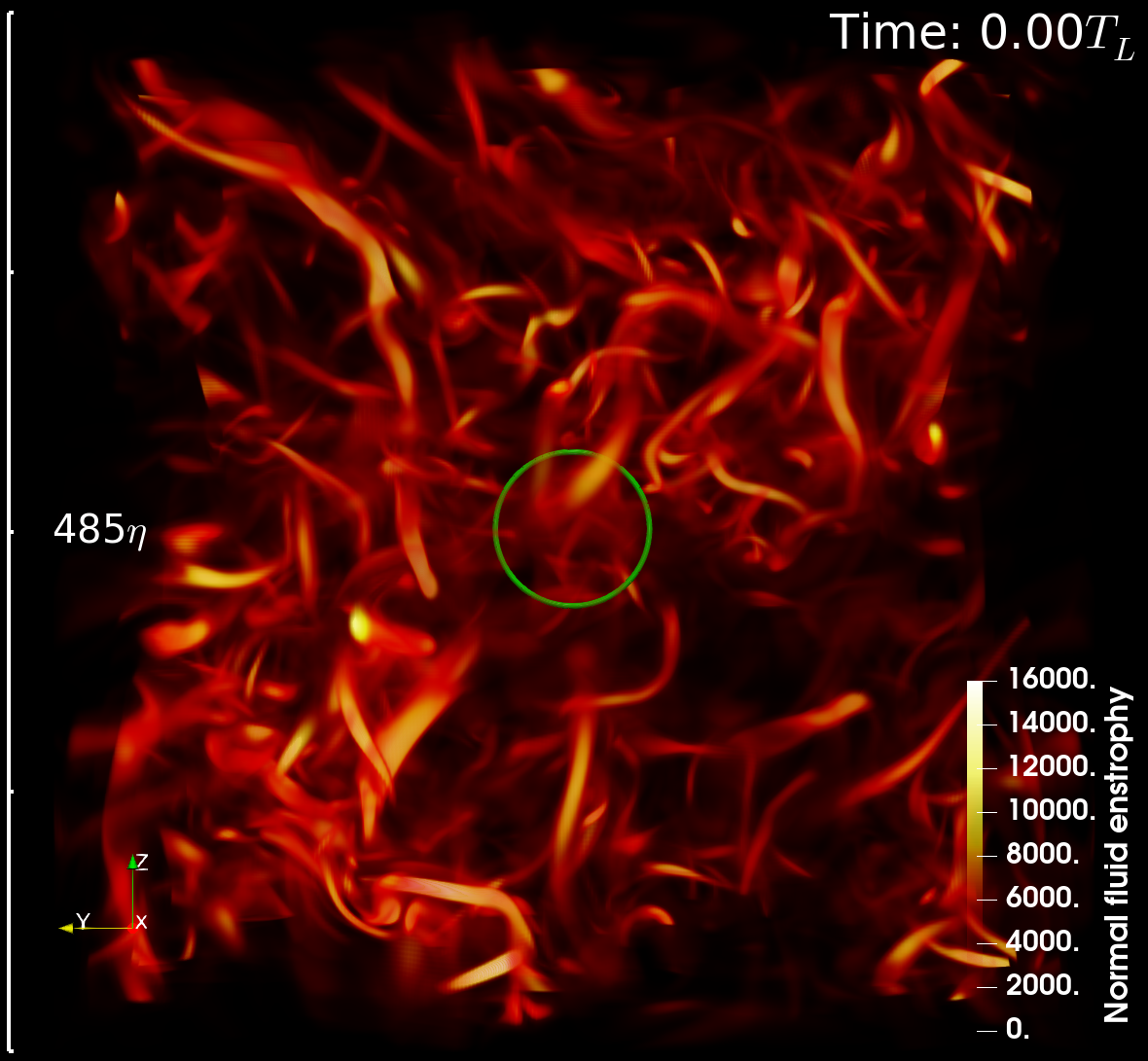}} 
    \subfigure{\includegraphics[width=0.49\textwidth]{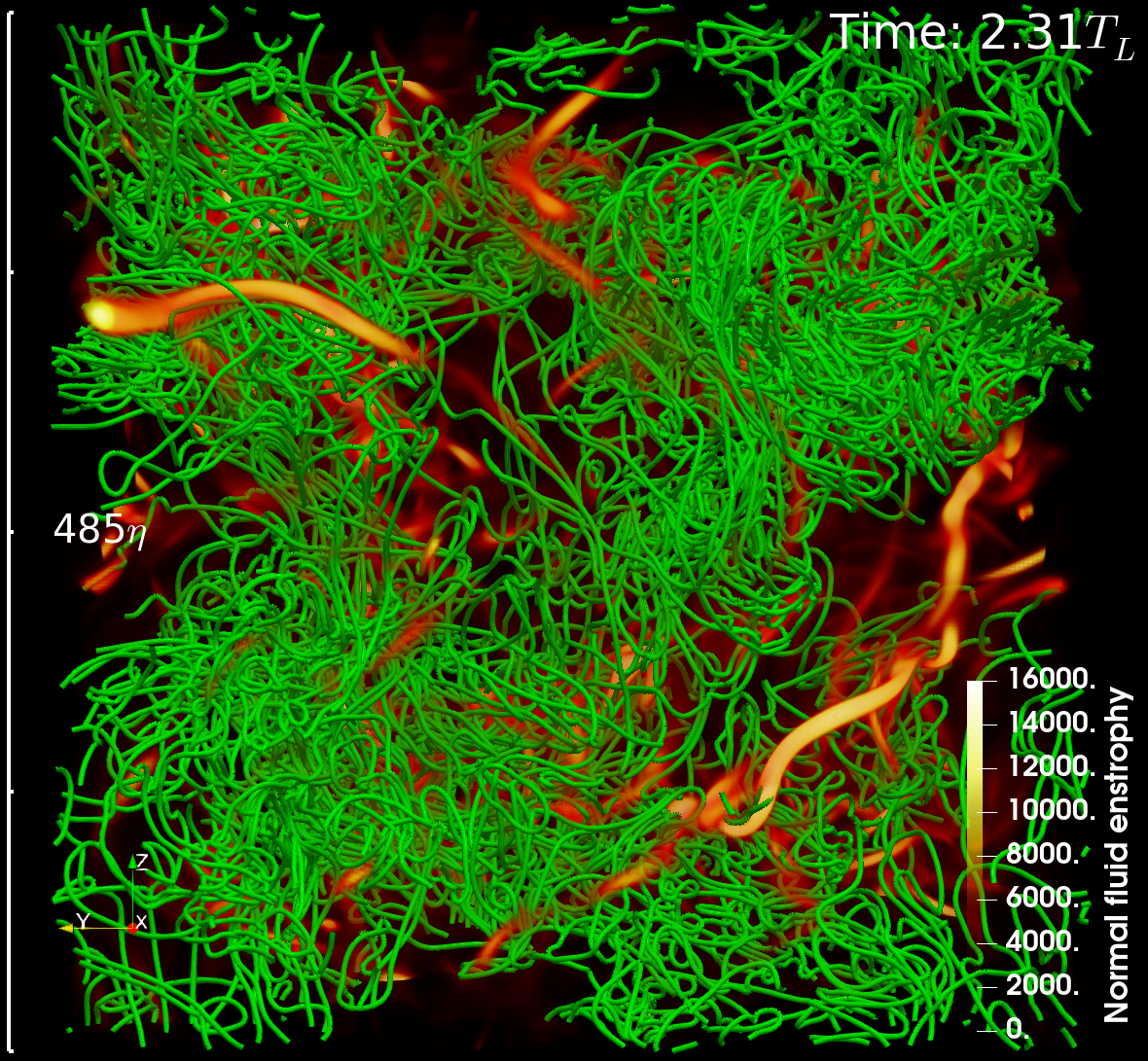}} \\
  \vspace*{-1mm}
    \subfigure{\includegraphics[width=0.49\textwidth]{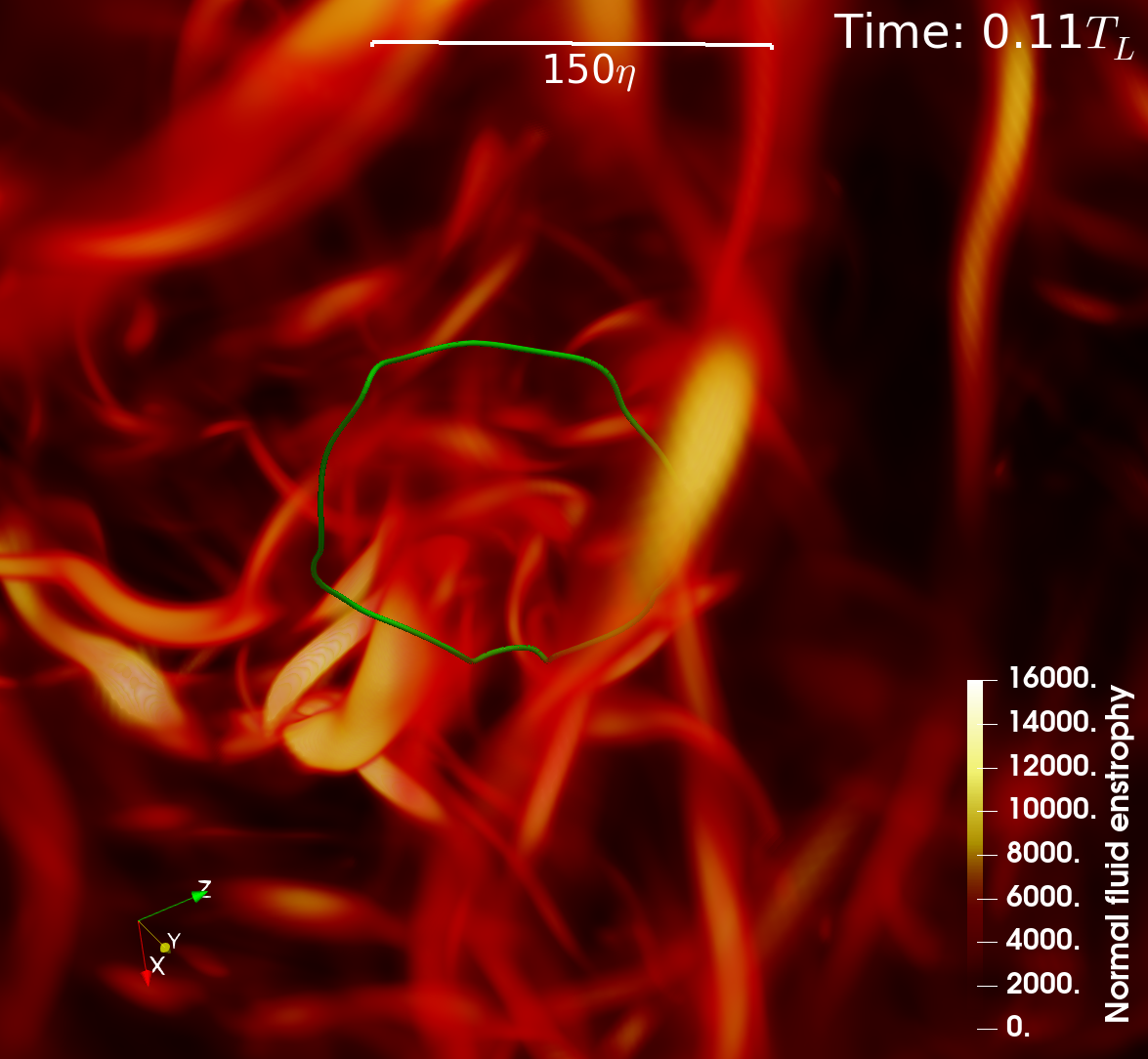}} 
    \subfigure{\includegraphics[width=0.49\textwidth]{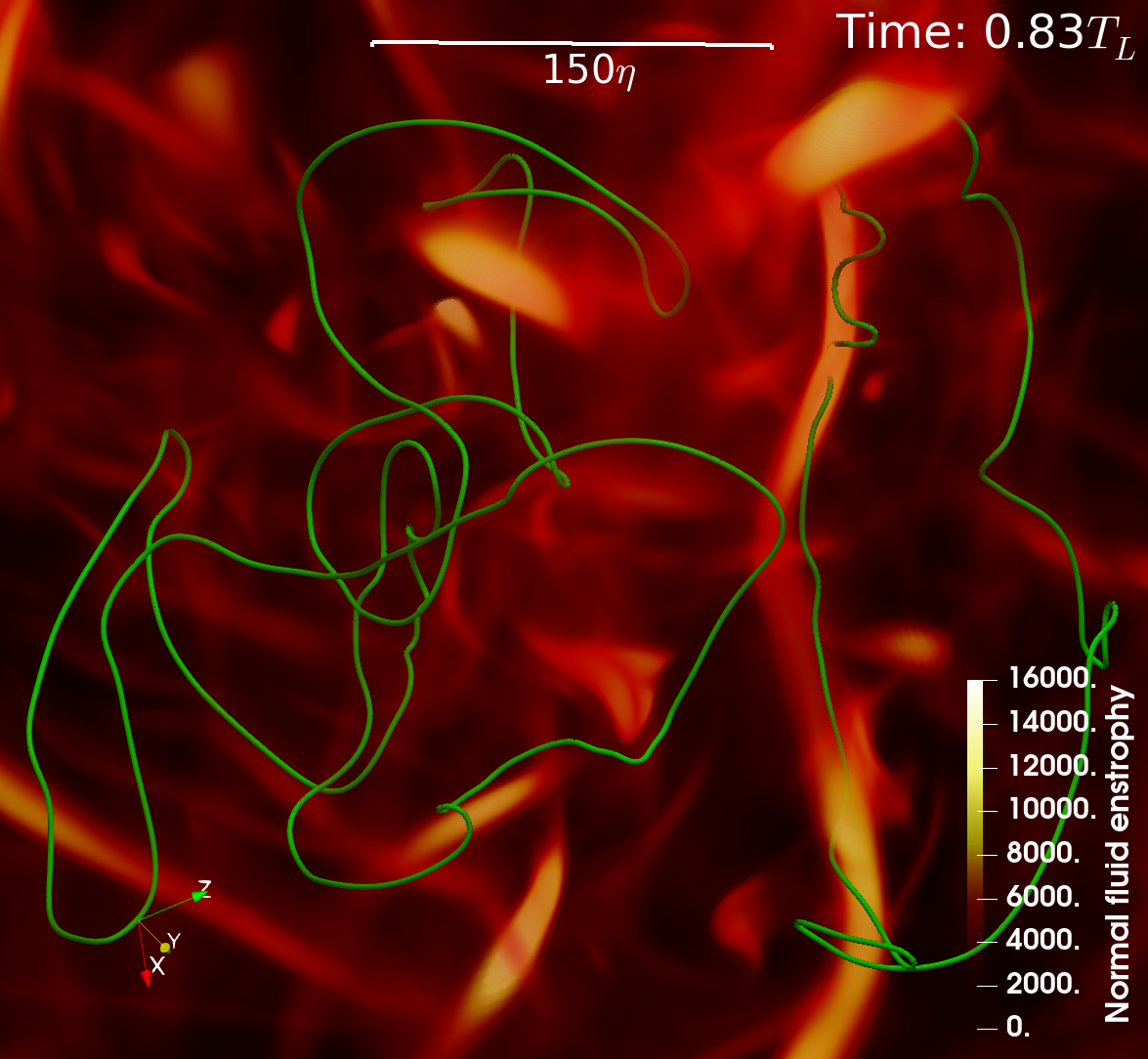}} 
    \caption{(Colour online). Evolution of a superfluid vortex ring 
in turbulent normal fluid. 
The superfluid vortex ring is visualised
by the green line. The normal fluid enstrophy is displayed by the 
yellow-orange-black rendering. The top row displays the full box at the initial (left) and the final (right) time of the simulation. The bottom row, show zoom close to the vortex ring at two early times. The figures are taken at 
$t=0$ (top left), 
$t=2.31T_L$ (top right), 
$t=0.11T_L$ (bottom left), and 
$t=0.83T_L$ (bottom right), 
where $T_L$ is the large eddy-turnover time of the normal fluid. The solid white lines indicate the size, expressed in units of the Kolmogorov length $\eta$ ,of the structures in the flow.
\label{Fig:VizRingInTurb}}
%\end{centering}
\end{figure}
We clearly observe the turbulent fluctuations of the normal fluid 
displayed by the reddish rendering of the normal fluid enstrophy. 
As time evolves, the turbulent fluctuations destabilise the superfluid
vortex ring, inducing large amplitude Kelvin waves (bottom left panel);
After a couple of large-eddy-turnover times, the vortex ring
self-reconnects, forming vortex loops which undergo further
reconnections, creating a turbulent superfluid
vortex tangle in equilibrium with the turbulent normal fluid.
It is apparent from Fig.~\ref{Fig:VizRingInTurb} 
that normal fluid fluctuations are responsible for an increase of the 
total vortex length. In Fig.\ref{Fig:VortexLengthRingInTurb} we display 
the temporal dependence of the total superfluid vortex length:
the large amount of stretching is evident. The right panel shows
that initially the vortex
length does not increase exponentially, as predicted for the
Donnelly-Glaberson instability \cite{Tsubota2003} 
for a uniform normal flow, but with approximate power-law character.
The figure also shows that, 
remarkably, there is a change of behaviour 
at times of the order of one $T_L$, suggesting that 
normal fluid fluctuations may play an important role 
in quantum turbulence. This is an interesting question 
with important implications for current quantum turbulence 
experiments.  It is natural to expect that the evolution of the 
vortex length should depend on temperature, but
be independent \cite{schwarz-1988} of the initial 
vortex configuration.
Clearly, an accurate and physically sound solver of the fully-coupled 
dynamics of normal fluid and superfluid like the solver that we have presented
here is an important tool for the next studies of
quantum turbulence.

\end{enumerate}

%%% CONCLUSION %%%%%%%%%%%%%%%

\section{Conclusions}
\label{sec:concl}

We have presented a novel algorithm, named \textit{FOUCAULT}, for the 
numerical simulations of quantum turbulence in helium~II
at nonzero temperatures. 
The main features of our approach are
(i) the parallelised, efficient, pseudo-spectral code, capable of 
distributing the calculation amongst distinct computational
cluster codes, and (ii) the forcing scheme employed for the normal fluid. 
These features allow the resolution
a wider range of length scales compared to previous studies, 
from large quasi-classical scales to small quantum length scales smaller
than the average inter-vortex distance.
This is of fundamental importance in order to 
investigate the character of quantum turbulence shared with its
classical counterpart, as well as
the features which are specific to turbulent superfluid flows.
%and to achieve a statistically stationary turbulent state,
%not yet achieved in literature.    

From the point of view of the physical modelling,
a novel feature of our approach is the local computation of the friction 
force, stemming from classical creeping flow analysis, which 
(unlike previous work) is
independent of the numerical discretisation on the vortex lines.
In addition, our approach implements an exact regularisation 
of the friction force based on the small-scale viscous 
diffusion of the disturbances generated by vortex lines moving with
respect to the normal fluid.
%Overall, the numerical algorithm contains thus
%significative innovations with respect to numerical codes previously described in literature.

After a detailed description of the distinct features of our algorithm, 
we have applied it
%validated it by calculating the normal fluid structures 
to two problems: the motion of a superfluid vortex ring in an initially
stationary normal fluid and in a turbulent normal fluid.
In the first problem we have recovered the same qualitative
features (the double normal ring structure and wake) observed in
previous work; in the second we have demonstrated a new effect - the
vortex stretching induced by turbulent fluctuations.

In conclusion, our novel algorithm paves the way for a new series of 
investigations of quantum turbulence in helium~II at finite 
temperatures, particularly at length scales smaller than the average inter-vortex
spacing in problems where the singular and quantised nature of the superfluid 
vortices is expected to play a fundamental role.

\section*{Acknowledgements}
This work was supported by the Royal Society International Exchange Grant n. IES \textbackslash R2\textbackslash 181176. 
L.G and C.F.B. also acknowledge the support of the Engineering and Physical Sciences Research Council 
(Grant No. EP/R005192/1). G.K. was also supported by the Agence Nationale de la Recherche through the 
project GIANTE ANR-18-CE30-0020-01. Computations were carried out on the Mésocentre SIGAMM hosted at the 
Observatoire de la Côte d’Azur and on the HPC Rocket Cluster at Newcastle University
The authors also acknowledge the University of Washington, Institute of Nuclear Theory program entitled 
\textit{INT-19-1a Quantum Turbulence: Cold Atoms, Heavy Ions, and Neutron Stars}, where part of the research was performed.

\end{document}